\documentclass[12pt]{article}
\usepackage{fancyhdr}

\usepackage{mathrsfs}
\usepackage[T1]{fontenc}
\usepackage{mathpazo}
\usepackage{setspace}
\usepackage{amsfonts}
\usepackage{amssymb}
\usepackage{amsmath}
\usepackage{epsfig}
\usepackage{latexsym}
\usepackage{color}
\usepackage{graphicx}
\usepackage{nicefrac}
\usepackage[latin1]{inputenc}
\usepackage{slashed}
\usepackage{multirow}
\usepackage{comment}
\usepackage{soul}
\usepackage{hyperref}
\usepackage[nosort]{cite}
\usepackage{datetime}

\usepackage{braket}

\usepackage[titletoc]{appendix}

\def\hybrid{\topmargin -20pt    \oddsidemargin 0pt
        \headheight 0pt \headsep 0pt
        \textwidth 6.25in       
        \textheight 9.25in       
        \marginparwidth .875in
        \parskip 5pt plus 1pt   \jot = 1.5ex}

\hybrid

\def\baselinestretch{1.2}

\catcode`\@=11

\def\marginnote#1{}
\newcount\hour
\newcount\minute
\newtoks\amorpm
\hour=\time\divide\hour by60
\minute=\time{\multiply\hour by60 \global\advance\minute by-\hour}
\edef\standardtime{{\ifnum\hour<12 \global\amorpm={am}%
        \else\global\amorpm={pm}\advance\hour by-12 \fi
        \ifnum\hour=0 \hour=12 \fi
        \number\hour:\ifnum\minute<10 0\fi\number\minute\the\amorpm}}
\edef\militarytime{\number\hour:\ifnum\minute<10 0\fi\number\minute}

\def\draftlabel#1{{\@bsphack\if@filesw {\let\thepage\relax
   \xdef\@gtempa{\write\@auxout{\string
      \newlabel{#1}{{\@currentlabel}{\thepage}}}}}\@gtempa
   \if@nobreak \ifvmode\nobreak\fi\fi\fi\@esphack}
        \gdef\@eqnlabel{#1}}
\def\@eqnlabel{}
\def\@vacuum{}
\def\draftmarginnote#1{\marginpar{\raggedright\scriptsize\tt#1}}

\def\draft{\oddsidemargin -.5truein
        \def\@oddfoot{\sl preliminary draft \hfil
        \rm\thepage\hfil\sl\today\quad\militarytime}
        \let\@evenfoot\@oddfoot \overfullrule 3pt
        \let\label=\draftlabel
        \let\marginnote=\draftmarginnote
   \def\@eqnnum{(\theequation)\rlap{\kern\marginparsep\tt\@eqnlabel}%
\global\let\@eqnlabel\@vacuum}  }

\def\preprint{\twocolumn\sloppy\flushbottom\parindent 2em
        \leftmargini 2em\leftmarginv .5em\leftmarginvi .5em
        \oddsidemargin -.5in    \evensidemargin -.5in
        \columnsep .4in \footheight 0pt
        \textwidth 10.in        \topmargin  -.4in
        \headheight 12pt \topskip .4in
        \textheight 6.9in \footskip 0pt
        \def\@oddhead{\thepage\hfil\addtocounter{page}{1}\thepage}
        \let\@evenhead\@oddhead \def\@oddfoot{} \def\@evenfoot{} }

\def\numberbysection{\@addtoreset{equation}{section}
        \def\theequation{\thesection.\arabic{equation}}}

\def\underline#1{\relax\ifmmode\@@underline#1\else
        $\@@underline{\hbox{#1}}$\relax\fi}

\def\titlepage{\@restonecolfalse\if@twocolumn\@restonecoltrue\onecolumn
     \else \newpage \fi \thispagestyle{empty}\c@page\z@
        \def\thefootnote{\fnsymbol{footnote}} }

\def\endtitlepage{\if@restonecol\twocolumn \else \newpage \fi
        \def\thefootnote{\arabic{footnote}}
        \setcounter{footnote}{0}}  

\catcode`@=12
\relax

\def\figcap{\section*{Figure Captions\markboth
        {FIGURECAPTIONS}{FIGURECAPTIONS}}\list
        {Figure \arabic{enumi}:\hfill}{\settowidth\labelwidth{Figure
999:}
        \leftmargin\labelwidth
        \advance\leftmargin\labelsep\usecounter{enumi}}}
 \relax
\def\tablecap{\section*{Table Captions\markboth
        {TABLECAPTIONS}{TABLECAPTIONS}}\list
        {Table \arabic{enumi}:\hfill}{\settowidth\labelwidth{Table
999:}
        \leftmargin\labelwidth
        \advance\leftmargin\labelsep\usecounter{enumi}}}
 \relax
\def\reflist{\section*{References\markboth
        {REFLIST}{REFLIST}}\list
        {[\arabic{enumi}]\hfill}{\settowidth\labelwidth{[999]}
        \leftmargin\labelwidth
        \advance\leftmargin\labelsep\usecounter{enumi}}}
 \relax
\makeatletter
\newcounter{pubctr}
\def\publist{\@ifnextchar[{\@publist}{\@@publist}}
\def\@publist[#1]{\list
        {[\arabic{pubctr}]\hfill}{\settowidth\labelwidth{[999]}
        \leftmargin\labelwidth
        \advance\leftmargin\labelsep
        \@nmbrlisttrue\def\@listctr{pubctr}
        \setcounter{pubctr}{#1}\addtocounter{pubctr}{-1}}}
\def\@@publist{\list
        {[\arabic{pubctr}]\hfill}{\settowidth\labelwidth{[999]}
        \leftmargin\labelwidth
        \advance\leftmargin\labelsep
        \@nmbrlisttrue\def\@listctr{pubctr}}}
 \relax
\makeatother
\newskip\humongous \humongous=0pt plus 1000pt minus 1000pt

\newif\ifdtup

\relax

\def\be{\begin{equation}}
\def\ee{\end{equation}}
\def\ba{\begin{eqnarray}}
\def\ea{\end{eqnarray}}

\def\del{\partial}

\def\k{\kappa}

\def\a{\alpha}

\def\b{\beta}

\def\g{\gamma}

\def\d{\delta}
\def\D{\Delta}
\def\e{\epsilon}

\def\p{\pi}

\def\th{\theta}

\def\m{\mu}

\def\l{\lambda}
\def\L{\Lambda}
\def\s{\sigma}

\def\cL{{\cal L}}

\def\no{\noindent}

\def\qq{\qquad}

\def\IR{\relax{\rm I\kern-.18em R}}

\def \z { {\bar z} }

\def \J {{\bar J} }

\def \ha {{1\over 2}}

\def \ov {\over}

\def\IR{\relax{\rm I\kern-.18em R}}
\def\IL{\relax{\rm I\kern-.18em L}}

\def\inv{^{\raise.15ex\hbox{${\scriptscriptstyle -}$}\kern-.05em 1}}

\def\cL{{\cal L}}

\def\Tr{{\rm Tr}}

\begin{document}

\renewcommand{\theequation}{\thesection.\arabic{equation}}
\csname @addtoreset\endcsname{equation}{section}

\newcommand{\beq}{\begin{equation}}
\newcommand{\eeq}[1]{\label{#1}\end{equation}}
\newcommand{\ber}{\begin{equation}}
\newcommand{\eer}[1]{\label{#1}\end{equation}}
\newcommand{\eqn}[1]{(\ref{#1})}
\begin{titlepage}
\begin{center}


${}$
\vskip .2 in

\vskip .4cm

{\large\bf
Exact results from the geometry of couplings and the effective action
}

\vskip 0.4in

{\bf George Georgiou}, {\bf Pantelis Panopoulos},\\
\vskip .08cm
 {\bf Eftychia Sagkrioti}\hskip .2 cm and \hskip .15 cm {\bf Konstantinos Sfetsos}
\vskip 0.16in

 {\em
Department of Nuclear and Particle Physics,\\
Faculty of Physics, National and Kapodistrian University of Athens,\\
Athens 15784, Greece\\
}

\vskip 0.12in

{\footnotesize \texttt {ggeo,ppanopoulos,esagkrioti,ksfetsos@phys.uoa.gr}}


\vskip .5in
\end{center}

\centerline{\bf Abstract}

\no
We invent a method that exploits the geometry in the space of couplings and the known all-loop effective action, in order to calculate the exact in the couplings anomalous dimensions of composite operators for a wide class of integrable $\s$-models.
These involve both self and mutually interacting current algebra theories.
We work out the details for important classes of such operators. In particular, we consider the operators built solely from an arbitrary number of currents of the same chirality, the composite operators which factorize into a chiral and an anti-chiral part, as well as those made up
of three currents of mixed chirality. Remarkably enough, the anomalous dimensions of the former two sets of operators turn out to vanish.
In our approach, loop computations are completely avoided.

\vskip .4in
\noindent
\end{titlepage}
\vfill
\eject

\newpage

\tableofcontents

\noindent

\def\baselinestretch{1.2}
\baselineskip 20 pt
\noindent


\setcounter{equation}{0}

\section{Introduction }

Two quantities of particular importance exist in any quantum field theory (QFT). Namely, the $\b$-functions governing the running of the coupling constants with the energy scale and the anomalous dimensions of fundamental and composite operators encoding their scaling  properties. Usually, both of them are determined perturbatively order by order.

In this work, we will develop a method which allows one to calculate {\it exact} expressions for both the $\b$-functions and the anomalous dimensions of large classes of operators, including composite ones, at one go. The power of our method will be exhibited by considering  a certain class of integrable two-dimensional $\sigma$-models. These models can be obtained by deforming conformal field theories (CFTs) of the WZW type with bilinear current-current operators. The all-loop, in the deformation parameters but at large WZW levels, effective action of these models was obtained through a gauging procedure initiated in \cite{Sfetsos:2013wia} and further developed and exploited in \cite{Georgiou:2016urf,Georgiou:2017jfi,Georgiou:2018hpd,Georgiou:2018gpe} for the cases of equal and unequal levels where mutual and/or self interactions are present. The possibility of using non-trivial automorphisms in the context of single $\l$-deformations  was put forward in  \cite{Driezen:2019ykp}.
These models are collectively called $\l$-deformations as they are characterized by a square matrix
$\l$ with dimensionality equal to that of the underlying group structure.
In addition, {\it exact} expressions for three-point correlators of currents and primary fields were calculated for the prototype $\l$-deformed models \cite{Sfetsos:2013wia} possessing a single level in \cite{Georgiou:2016iom} and for the case of two unequal levels of \cite{Georgiou:2017jfi} in \cite{Georgiou:2016zyo}.
The aforementioned models are particularly attractive because they possess certain non-perturbative symmetries in the space of couplings \cite{Kutasov:1989aw,Itsios:2014lca,Sfetsos:2014jfa,Georgiou:2017jfi,Georgiou:2018hpd,Georgiou:2018gpe}. As a result, one can exploit these non-perturbative symmetries by combining them with low-order perturbative calculations in order to derive {\it exact} expressions for the $\b$-functions and the anomalous dimensions of current and primary operators \cite{Georgiou:2015nka,Georgiou:2016iom,Georgiou:2017aei}.\footnote{The $\b$-functions can also be computed by using a variety  of methods. One can either re-sum the series of  conformal perturbation  theory \cite{Kutasov:1989dt,Gerganov:2000mt,LeClair:2001yp} or exploit the effective action and use a variant of the background field method  \cite{Appadu:2015nfa,Georgiou:2017oly,Sagkrioti:2018rwg}. An independent method  is to use the well-known expressions for the  $\b$-functions for the metric and the antisymmetric tensors fields in non-linear $\s$-models \cite{honer}
which was done in \cite{Itsios:2014lca,Sfetsos:2014jfa,Georgiou:2017jfi,Sagkrioti:2018rwg}.}
This approach is particularly elegant and effective but it has the drawback that in practice it is difficult to apply for composite operators
made up of many fundamental ones since the difficulty of perturbative calculations increases enormously with the
length of the operator.  A less apparent drawback is that the form of bare operator, i.e. when the deformation is switched off,
changes when this is turned on. Both of these drawbacks are rectified in the present work.

Remarkably, in these $\l$-deformed $\s$-models the computation of Zamolodchikov's $C$-function \cite{Zamolodchikov:1986gt}, as an exact function of the deformation
parameters and to leading order in the large $k$-expansion, is possible.
This was first performed in \cite{c-function:2018} for the isotropic cases and further generalized for anisotropic $\l$-deformations in \cite{Sagkrioti:2018abh}.

One of the virtues of the aforementioned constructions \cite{Sfetsos:2013wia,Georgiou:2017jfi,Georgiou:2018hpd,Georgiou:2018gpe} is that the deformed models are integrable for
specific forms of the deformation matrix. Besides the case for isotropic such matrices, i.e. when the matrix
$\l$ is proportional to the identity,
integrability holds for subclasses of anisotropic models as well. In particular, for the $\l$-deformed $SU(2)$ based models in \cite{Sfetsos:2014lla,Sfetsos:2015nya}, as well as for certain subclasses of those  in \cite{Georgiou:2018hpd,Georgiou:2018gpe}.\footnote{For the case of the isotropic deformation based on the $SU(2)$ group integrability has been demonstated in \cite{Balog:1993es}.} Integrable deformations based on cosets, symmetric and semi-symmetric spaces were also constructed and studied in \cite{Sfetsos:2013wia,Sfetsos:2017sep}, \cite{Hollowood:2014rla} and \cite{Hollowood:2014qma}, respectively.
In certain cases, deformed models of low dimensionality were promoted to solutions of type-IIA or type-IIB supergravity \cite{Sfetsos:2014cea,Demulder:2015lva,Borsato:2016zcf,Chervonyi:2016ajp,Borsato:2016ose}.

Another interesting feature of the $\l$-deformations is their relation to the so-called  $\eta$-deformations with the latter being introduced in \cite{Klimcik:2002zj,Klimcik:2008eq,Klimcik:2014}  for group spaces
and in \cite{Delduc:2013fga,Delduc:2013qra,Arutyunov:2013ega} for coset spaces. This relation is realized by the action of a Poisson-Lie T-duality \cite{KS95a,Sfetsos:1999zm} and appropriate analytic continuations which for group and coset spaces was discussed in  \cite{Vicedo:2015pna,Hoare:2015gda} and \cite{Sfetsos:2015nya,Klimcik:2015gba,Klimcik:2016rov,Hoare:2018ebg}, respectively.
In parallel developments, the dynamics of scalar fields in certain $\l$-deformed geometries based on
coset CFTs has been discussed in \cite{Lunin:2018vsn} while the realization of $\l$-deformations as theories living on the boundary of Chern-Simons theories was discussed
in \cite{Schmidtt:2017ngw}. Finally,  D-branes  in the context of single $\lambda$-deformations were studied in
\cite{Driezen:2018glg,Driezen:2019ykp}.

\no
The plan of the paper is as follows:  In section 2, we introduce the essential features of our method and use it to derive the anomalous dimension of the fundamental current in complete agreement with previous calculations \cite{Georgiou:2015nka}. In section 3, we employ our method to calculate the anomalous dimensions of composite operators build from chiral and/or anti-chiral holomorphic currents. As two explicit examples we consider operators built solely from an arbitrary number of chiral currents as well as operators factorizing into chiral and anti-chiral parts.
Surprisingly enough, their anomalous dimensions turn out to be zero to leading order in the large $k$-expansion. Our third example concerns the fully symmetric operator composed from two chiral and one anti-chiral current. The result  for its anomalous dimension is the same as the dimension of the operator $J_+J_-$ that drives the model off conformality. In all cases the expressions for the anomalous dimensions respect the non-perturbative symmetries of the model \cite{Kutasov:1989aw}.
In section 4, we focus on the model constructed in \cite{Georgiou:2017jfi}. As in section 3, the anomalous dimensions of purely chiral or anti-chiral current operators build from an arbitrary number of currents are also zero. The same is true
for operators that factrorize into chiral and anti-chiral parts. Similarly,
the dimension of the mixed operator matches again the dimension of the operator that drives the model off conformality. In section 5, we consider the model where both self and mutual current-current interactions are present \cite{Georgiou:2018hpd}.
Using the geometry in the space of couplings and the exact $\b$- functions of this model, we calculate the exact in the deformation parameters anomalous dimensions of the operators which perturb the CFT, as well as those of the single currents. In section 6, we present a number of perturbative calculations which are in agreement with the exact results obtained in the previous sections. Finally, in section 7 we draw our conclusions. We have also written four appendices containing
technical and computational details.

\section{The anomalous dimension of the single current}

In this section we will explain the essential features of our method. Focusing on the single $\l$-deformed $\s$-models
\cite{Sfetsos:2013wia} we will show how to compute the anomalous dimensions of the fundamental currents of the model. The method is based on a
convenient modification of the gauging
procedure of \cite{Sfetsos:2013wia} in conjunction with geometrical data defined in the space of couplings of the corresponding two-dimensional field theories. Having the essential and conceptual aspects of the method
under control we will extend the construction to include composite operators of currents in the next two sections.

\no
We start with the sum of the WZW model action $S_{k}(g)$ at level $k$ for a
group element $g$ in a group $G$  \cite{Witten:1983ar}, the principal chiral model (PCM) action \cite{Polyakov:1975rr}
for the group element $\tilde g\in G$
with overall coupling $\k^2$ and a term containing the chiral current of the original WZW model. Specifically this action reads
\be
\label{oractt}
 S_{k,\k^2,s}(g,\tilde g)= S_{k}(g)
-{\k^2\ov \pi}\int d^2\s\ \Tr\big(\tilde g^{-1}\del_+\tilde g \tilde g^{-1} \del_-\tilde g  \big)
+  {k\ov \pi}\int d^2\s\ \Tr\big(s\tilde g^{-1}\del_+\tilde g \big)\  ,
\ee
where the last new term has coupling matrix $s=s^a t_a$ and the overall scaling $k$ has been introduced for later convenience.
This extra term is auxiliary and the reason for adding it will become progressively apparent in this section. It will eventually enable us to compute the exact anomalous dimension of the fundamental current in the $\l$-deformed theory of \cite{Sfetsos:2013wia} surpassing perturbation theory.

All matrices are expanded using as a basis representation matrices $t_a$
obeying $[t_a,t_b]=i f_{abc}t_c$ and normalized as $\Tr(t_a t_b)=\d_{ab}$.                                                                                                                                                                                             As in \cite{Sfetsos:2013wia} we gauge the global symmetry acting as
$g\to \L^{-1} g\L$ and $\tilde g\to \L^{-1} \tilde g$. The corresponding gauge invariant action is
\be
\label{gauacc}
\begin{split}
& S_{k,E}(g,\tilde g, A_\pm)  = S_{k}(g,A_\pm)
-{\k^2\ov \pi}\int d^2\s\ \Tr\big( \tilde g^{-1}D_+\tilde g  \tilde g^{-1} D_-\tilde g\big)
\\
& \qq\qq\qq\quad
+  {k\ov \pi}\int d^2\s\ \Tr\big(s\tilde g^{-1}D_+\tilde g \big)\ ,
\end{split}
\ee
where $D_\pm \tilde g= (\del_\pm -A_\pm) \tilde g$ are the covariant derivatives.
The first term is the standard gauged WZW action \cite{gwzwac}
\be
\begin{split}
&  S_{k}(g,A_\pm) = S_{k}(g)
+{k\ov \pi} \int d^2\s \ \Tr \big(A_- \del_+ g g^{-1}   - A_+ g^{-1} \del_- g
\\
& \qq\qq\quad + A_- g A_+ g^{-1}-A_-A_+\big)\ .
\end{split}
\ee
Fixing the gauge in \eqn{gauacc} by choosing $\tilde g=\mathbb{1}$ one arrives at the following action
\be
\begin{split}
& S_{k,\l,s}(g, A_\pm) = S_{k}(g)
+{k\ov \pi} \int d^2\s \ \Tr \big(A_- \del_+ g g^{-1}   - A_+ g^{-1} \del_- g+ A_- g A_+ g^{-1}\big) \\
&\qq\qq\qq - {k \ov \pi}  \int d^2\s\ \Tr\big(\l^{-1} A_+ A_- + s A_+\big)\ ,
\label{gaufix}
\end{split}
\ee
where
\be
\l^{-1}= 1+{\k^2 \ov k}\ .
\ee
We are interested in the equations of motion of this action.
Varying \eqn{gaufix} with respect
to $A_\mp$ we find the constraints
\be
\label{dggd}
\begin{split}
&
D_+ g\, g^{-1} +  (1-\l^{-1}) A_+ =0 \ ,
\\
&
g^{-1} D_- g  - (1-\l^{-1})  A_- + s =0 \ ,
\end{split}
\ee
where the covariant derivatives acting on $g$ are now defined as $D_\pm g= \del_\pm g -[A_\pm,g]$.
Variation with respect to the group element $g$ gives
\be
\label{eqg1g2}
D_ -(D_+ g g^{-1})= F_{+-} \quad \Longleftrightarrow\quad D_+(g^{-1}D_- g )
= F_{+-}\ ,
\ee
where the field strength is as usual
\be
F_{+-}=\del_+ A_- - \del_- A_+ - [A_+,A_-]\ .
\ee
Substituting \eqn{dggd} into \eqn{eqg1g2} we obtain that
\be
\begin{split}
\label{eomAinitial1}
& \del_- A_+  - \l \del_+A_- + [A_+,  A_-]=0 \ ,
\\
& \del_+ A_- -\l \del_-  A_+ -  [  A_+,A_-] -\l [A_+, s]=0 \ .
\end{split}
\ee
We may use the constraints \eqn{dggd} to solve for the gauge fields and upon substitution back into the action
\eqn{gaufix} arrive at a $\s$-model action.
The result is nothing by the $\l$-deformed $\s$-model action corresponding to the $\l$-deformed models in the isotropic case,
plus a term  linear in  $s$. Specifically, we find that
\be
\label{apam1}
A_+= i \big(\l^{-1}\mathbb{1} - D)^{-1} J_+\ , \qq A_- =-  \big(\l^{-1}\mathbb{1} - D^T)^{-1} (iJ_- +s)\ ,
\ee
where
\be
J_+ =- i \del_+ g g^{-1} \ ,\quad J_- = - i g^{-1} \del_- g\ ,\quad D_{ab}=\Tr(t^a g t^b g^{-1})\ .
\ee
Then the action becomes
\be
\label{djkg11}
S= S_k(g) + {k\ov \pi} \int d^2\s\  J_+ (\l^{-1}\mathbb{1} - D^T)^{-1} J_-
- {k\ov \pi} \int d^2 \s\ \Tr(s A_+) ,
\ee
Obviously, for $s=0$ the original $\l$-deformed theory \cite{Sfetsos:2013wia} is recovered. We are interested in computing the anomalous dimensions of the current $J_+^a$ exactly in $\l$ and to leading order in $k$ at the limit $s=0$, that is
for the original $\l$-deformed  theory. This limit has to be consistent with the $\b$-function equations, which is the case as we shall see below. Note that the current $J_+$ is ``dressed'' and replaced by $A_+$ as it can be seen from the corresponding
last term in \eqn{djkg11}. Obviously for $\l\to 0$ we have that $A_+\sim J_+$.

\subsection{The RG flow equations}

Next we compute the $\b$-function equations for the couplings $\l$ and $s$.
We will follow the background-type method initiated for the $\l$-deformed models in\cite{Appadu:2015nfa}
and applied in full generality in \cite{Sagkrioti:2018rwg}.
We choose a particular configuration of the group elements in order to compute the running of couplings. In particular,
$g = e^{\s^+ \th_+ + \s^- \th_-}$, where the  matrices  $\th_\pm$
are constant and commuting. Then we have that $J_{\pm}=-i \th_\pm$ and that the matrix $D=\mathbb{1}$.
Then, from \eqn{apam1} the background gauge fields are
\be
\label{gg23}
A^{(0)}_+={\l\ov 1-\l} \th_+\ ,\qq  A^{(0)}_-=-{\l\ov 1-\l}(\th_- +s )\ .
\ee
Note that, indeed the above solve the classical equations  \eqn{eomAinitial1}.
Then the Lagrangian density for the action \eqn{djkg11} reads
\be
\cL^{(0)}=-{k\ov 2 \pi}  \Big( {1+\l\ov 1-\l} \th_+\th_- + 2 {s \l\ov 1-\l}  \th_+  \Big)\ .
\label{gg233}
\ee
The next step is to consider the fluctuations of the gauge fields around \eqn{gg23} and let
\be
A_\pm = A^{(0)}_\pm + \d A_\pm \ ,\qq
(\tilde A_\pm^{(0)})_{ab} =i f_{abc} (A_\pm^{(0)})_c\ ,\qq \tilde s_{ab} =i f_{abc} s_c\ .
\ee
The linearized fluctuations for the equations of motion are  given by
\be
\begin{split}
&
-\big(\l  \del_+ + \tilde A_+^{(0)}\big)\d A_- + \big(\del_- +  \tilde A_-^{(0)}\big) \d A_+ = 0 \ ,
\\
&
\big(\del_+ +  \tilde A_+^{(0)}\big) \d A_- - \big(\l \del_- + \tilde A_-^{(0)} +\l \tilde s \big)\d A_+  = 0\ .
\end{split}
\ee
These can be cast in the following form
\be
\label{dhatt}
\hat D \left(
         \begin{array}{c}
           \d A_-  \\
           \d A_+ \\
         \end{array}
       \right) = 0\ ,
\ee
where the operator $\hat D$ is first order in the worldsheet derivatives.
After the Euclidean analytic continuation and in the momentum space we, in the conventions of \cite{Georgiou:2018hpd}, replace $(\del_+,\del_-) $ by $ \ha (\bar p,p)\equiv (p_+,p_-)$.
Then we have that $ \hat D = \hat C + \hat F $, where
\be
\label{chatt}
\hat C = \left(
           \begin{array}{cc}
              -\l p_+ &  p_-  \\
              p_+ &  -\l p_-  \\
           \end{array}
         \right)\ , \qq
\hat F =\left(
           \begin{array}{cc}
          - \tilde A_+^{(0)} &  \tilde A_-^{(0)} \\
        \tilde A_+^{(0)} & -\tilde A_-^{(0)} - \l \tilde s \\
           \end{array}
         \right)\  .
\ee
The matrix $\hat C$ contains all momentum dependence.
Integrating out the fluctuations, gives the effective Lagrangian of our model
\be
-\cL_{\rm eff} = \cL^{(0)} + \int^\m {d^2 p\ov (2\pi)^2} \ln (\det \hat D)^{-1/2}\ .
\ee
This integral is logarithmically divergent with respect to the UV mass scale
$\m$ which is isolated by performing the large momentum expansion of the integrand
and keeping terms proportional to $\displaystyle {1\ov |p|^2}$, where $|p|^2=p\bar p$.
Since $\hat C$ grows with $|p|$ we use the fact that
\be
\ln (\det \hat D)  = \ln \det \hat C + \Tr (\hat C^{-1} \hat F) -\ha \Tr(\hat C^{-1} \hat F)^2 + \cdots  \ .
\ee
The last tern is the only one contributing to the aforementioned logarithmic divergence which is important for our purposes, thus obtaining
\be
\label{fhhreff}
-\cL_{\rm eff} = \cL^{(0)} + {1\ov 16 \pi^2} \int^\m d^2 p\, \Tr(\hat C^{-1} \hat F)^2 + \cdots \ .
\ee
Next we use the polar coordinates parametrization $p=re^{i\phi}$, $\bar p=re^{-i\phi}$ in which
the integration measure is  $d^2 p = r dr d\phi$ and evaluate $\Tr(\hat C^{-1} \hat F)^2$. The dependence on $r$ is of the form $1/r^2$ which upon integration
gives the necessary factor of $\ln \m$.
Then
\be
\label{hgghr}
\begin{split}
 & -\cL_{\rm eff} = \cL^{(0)} +{c_G  \ov 2 \pi} \ln \m^2
\left({A^{(0)}_+ A^{(0)}_-\ov (1+\l)^2}  + {s \l A^{(0)}_+ \ov (1-\l)(1+\l)^2} \right)
\\
& \phantom{xxxxx} =  \cL^{(0)} - {c_G  \ov 2 \pi} \ln \m^2 {\l^2\ov (1-\l^2)^2}  \th_+\th_-\ ,
\end{split}
\ee
were we used \eqn{gg23} for  the background solution for the gauge fields.
Also, $\Tr(\tilde A_+^{(0)} \tilde A_-^{(0)})= c_G (A_+^{(0)})^a
(A_-^{(0)})^a$ and $\Tr(\tilde A_+^{(0)} \tilde s)= c_G (A_+^{(0)})^a
s^a$, where $c_G$ is the eigenvalue of the quadratic Casimir  in the adjoint
representation defined as $f_{acd}f_{bcd}=c_G \d_{ab}$.
In the rest we drop the index in $s^a$ since the result for the $\b$-function and later for the anomalous will be independent from it.

\no
As usual in field theory, we demand that the action \eqn{hgghr}
is $\m$-independent, i.e. $\del_{\ln \m^2} \cL_{\rm eff}=0$.
For $k\gg 1$ this derivative acts only on the coupling constants in $\cL^{(0)}$.
Then, defining as usual
$\b^\l = \del_{\ln \m^2} \l$ and $\b^{s} = \del_{\ln \m^2} s$, we obtain that
\be
\label{systrg}
\b^\l  = -{c_G\ov 2k} {\l^2\ov (1+\l)^2} \ ,
\qq
\b^{s} ={c_G\ov 2 k} {s\l  \ov (1-\l) (1+\l)^2}\  .
\ee
For later convenience it is important  to find the running of the coupling\footnote{The proportionality constant below is just an $-i$. This redefinition is due to the fact that $s$
is purely imaginary in our conventions, i.e. see \eqn{oractt}.
Its inclusion or ommission does not affect the final result for the anomalous dimension.}
\be
\tilde \l \sim s \l \ ,
\ee
which will replace $s$ in our considerations. Using the above we obtain that
 \be
\b^{\tilde \l} ={c_G\ov 2 k} {\l^2\tilde \l  \ov (1-\l) (1+\l)^2}\  .
\ee

\subsection{The current anomalous dimension}

So far we have kept the couplings $\l$ and $\tilde \l$ finite. For small values for these couplings we have from \eqn{djkg11} that
\be
\label{djkg11p}
S= S_k(g) + {k\ov \pi} \int d^2\s \big(\l J^a_+ J^a_- + \tilde \l J_+^a \big) + \cdots \ .
\ee
Keeping the discussion general, instead of the single current perturbation, we consider a general perturbation with operators
\be
\l^i {\cal O}_i \  .
\ee
Each one of them has a classical dimension and the $\b$-functions for the couplings are denoted by $\b^i$.

\no
There is a metric $G_{ij}^{(0)}$ in the space of these couplings defined via the two-point function of the
${\cal O}_i$'s \cite{Kutasov:1989dt} with line element $ds^2= G_{ij}^{(0)} d\l^i d\l^j$.
Renormalizability and the Callan--Symanzik equation give
\be
\g_i{}^j =\nabla_i\b^j +  \nabla^j\b_i = \del_i \b^j + G^{(0)jm}\left( G_{in}^{(0)} \del_m \b^{n} + \b^n \del_n G_{im}^{(0)}\right)\ ,
\ee
where $\b_i = G_{ij}^{(0)} \b^j$.
Consider the case of two couplings $\l$ and $\tilde \l$. In general there is a mixing of the two operators even if this is absent at the conformal point.
Consequently, their anomalous dimension will be encoded in a matrix $\g_i{}^j$ with all four elements non-vanishing.
We assume that one of the couplings, say $\tilde\l$, can be consistently set to zero with the corresponding $\b^{\tilde \l}=0$.
In that limit we assume that the mixing vanishes as well.
Then, only the entries $\g_\l{}^\l$ and  $\g_{\tilde \l}{}^{\tilde \l}$ will be non-zero.
In our case the operators are ${\cal O}_1=J_+^a J_-^a$ and ${\cal O}_2=J_+^a$.
The latter breaks Lorentz invariance, so that an operator mixing between them  may occur. However, in
the limit $\tilde\l \to 0$ Lorentz invariance is restored and therefore mixing between operators of different chirality is non-existing.

\no
Quite generally, near $\tilde\l=0$ we assume for the metric in the coupling space the form
\be
\label{gggg}
G_{\l\l}^{(0)}(\l,\tilde \l) = g_{\l\l}^{(0)}(\l)+ {\cal O}(\tilde \l) \ ,
\quad G_{\tilde \l\tilde \l}^{(0)}(\l,\tilde \l)= g_{\tilde \l\tilde \l}^{(0)}(\l) + {\cal O}(\tilde \l) \ ,
\quad G_{\l\tilde \l}^{(0)}=  {\cal O}(\tilde \l)\
 \ee
 and that $\b^{\tilde\l}={\cal O}(\tilde \l)$.
 These are a consequence of the decoupling assumption at $\tilde\l=0$.
 Then, in the limit $\tilde \l\to 0$  we find using \eqn{gggg} that
\be
\g_{{\cal O}_1} = \g_{\l}{}^{\l}=2 \del_{\l}\b^{\l} + \b^\l  \del_\l \ln g^{(0)}_{\l \l}\
\label{JbJ}
\ee
and also that
\be
\g_{{\cal O}_2} = \g_{\tilde \l}{}^{\tilde \l}=2 \del_{\tilde\l}\b^{\tilde \l} + \b^\l  \del_\l \ln g^{(0)}_{\tilde \l \tilde \l}\ .
\label{singleJ}
\ee
Hence, in that limit we have the original $\l$-deformed theory with just the coupling $\l$, i.e.
\eqn{djkg11} with the last term absent. However, as a bonus we have the expression for
the anomalous dimension of the  operator ${\cal O}_2$ as well, which was our goal.
Note that, only the expressions for these metrics $g^{(0)}_{ii}$ at the strict $k\to \infty$ limit are needed since
 in \eqn{JbJ} and in \eqn{singleJ} the $\b$'s are already of ${\cal O}(1/k)$. In addition, even the overall constant in their specific expressions is immaterial.

\no
 Specifically, in our case we have by using \eqn{memtrr} that
 \be
 g_{\l \l}^{(0)} = {\dim G\ov (1-\l^2)^2}\ ,\qq  g_{\tilde \l\tilde \l}^{(0)} \sim {1\ov 1-\l^2}\ ,
 \ee
Then we first find using \eqn{JbJ} that
\be
\g_{J_+ J_-} = -{2 c_G\ov k} \frac{ \lambda (1-\lambda(1-\lambda))}{(1-\lambda)(1+\lambda)^3}\ ,
\label{jhjf}
\ee
which indeed was computed in \cite{Georgiou:2016iom} exploiting this geometric method.
In addition, calculating the right hand side of \eqn{singleJ} we find that
\be
\label{ansingg}
\g_{J_+}= {c_G\ov k} {\l^2 \ov (1-\l)(1+\l)^3}\ .
\ee
This result was firstly found in \cite{Georgiou:2016iom} using the symmetries in the coupling space
\be
\label{symmkl}
k\to -k \ ,\qq \l\to {1\ov \l}\
\ee
and leading order results  from perturbative methods.

\no
Next, we will apply the same method to compute the anomalous dimensions of more complicated
composite operators.
\section{Anomalous dimensions of general current composites }

In this section,  we will extend the formalism explained in the previous section in order to calculate the anomalous dimensions of general operators of the form
\be
\label{form-op}
{\cal O}^{(m,n)}
=S_{a_1\dots a_m;b_1\dots b_n}J_+^{a_1}\dots J_+^{a_m}  J_- ^{b_1}\dots  J_-^{b_n}\ .
\ee
By construction the overall tensor coefficient should be symmetric in the first $m$ indices, as well as in the last $n$ ones, separately. However, there is no symmetry property relating  the $a_i$'s and the $b_i$'s. This tensor can be decomposed
into irreducible representations of the group $G$. As in the case of the single current the above operator will be
dressed to an operator ${\cal O}^{(m,n)}_{\l}$ whose expression will be presented below.

\no
Our starting point is the action \eqn{gauacc} but with the $s$-term in the second line replaced by
\be
\label{anssww}
 {k s\ov \pi}
 \int d^2\s\   S_{a_1\dots a_m; b_1\dots b_n}(\tilde g^{-1}D_+\tilde g)^{a_1}\dots ( \tilde g^{-1}D_+\tilde g)^{a_m}( \tilde g^{-1}D_-\tilde g)^{b_1}\dots ( \tilde g^{-1}D_-\tilde g)^{b_n}\ ,
\ee
times a factor of $(-1)^{m+n+1}$ which we introduce so that subsequent expressions are as simple as possible.
This action is still gauge invariant and the gauge fixing $\tilde g=\mathbb{1}$ condition leads to
\be
\begin{split}
\label{action-gen}
& S_{k,\l,s}(g, A_\pm) = S_{k}(g)
+{k\ov \pi} \int d^2\s \ \Tr \big(A_- \del_+ g g^{-1}   - A_+ g^{-1} \del_- g+ A_- g A_+ g^{-1}\big) \\
&\qq\qq\qq - {k \ov \pi}  \int d^2\s\ \left(\l^{-1} \Tr\big( A_+ A_-\big) + s  {\cal A}^{(m,n)}_{+ -}\right)\ ,
\end{split}
\ee
where
\be
\label{amnpm}
{\cal A}^{(m,n)}_{+ -}
=S_{a_1\dots a_m ;b_1\dots b_n}A_+^{a_1}\dots A_+^{a_m}  A_- ^{b_1}\dots  A_-^{b_n}\ .
\ee

\no
The equations of motion for \eqn{action-gen} with respect to $A_-$ and $A_+$ are
\be
\label{con1}
\begin{split}
&
D_+ g\, g^{-1} =  (\l^{-1}-1) A_+ +  n s   {\cal A}^{(m, n')}_{+ -}\ ,
\\
&
g^{-1} D_- g =  - (\l^{-1}-1)  A_- -m s {\cal A}^{( m'\!, n)}_{+ -},
\end{split}
\ee
where we have defined the vectors ${\cal A}^{(m, n')}_{+ -}$ and ${\cal A}^{(m', n)}_{+ -}$ with components
\be
\label{defin1}
\begin{split}
&\big({\cal A}^{( m'\!, n)}_{+ -}\big)_{a}=S_{a\cdots a_{m-1} ;b_1\dots b_n}A_+^{a_1}\dots A_+^{a_{m-1}}  A_- ^{b_1}\dots A_-^{b_n}\ ,
\\
&\big({\cal A}^{( m,  n')}_{+ -}\big)_b=S_{a_1\dots a_m ;b \dots b_{n-1}}A_+^{a_1}\dots
A_+^{a_m}  A_- ^{b_1}\dots  A_-^{b_{n-1}}\ .
\end{split}
\ee
Hence, the prime indicates that one index is not contracted and is left free in the corresponding tensor coefficient.

\no
Varying the action with respect to the group element $g$ results into the same equation \eqn{eqg1g2} since the
$s$-term in \eqn{action-gen} does not depend on it.
Substituting the constraints \eqn{con1} into \eqn{eqg1g2} we obtain
\be
\begin{split}
\label{eomAinitial11}
& \l^{-1} \del_+  A_- -\del_-A_+
= \l^{-1}[ A_+, A_-] -m \, s \, D_+ {\cal A}^{( m'\!, n)}_{+ -} \ ,
\\
& \l^{-1} \del_-  A_+ -\del_+A_-
=-  \l^{-1}  [A_+,A_-] -n \, s \, D_- {\cal A}^{( m,  n')}_{+ -} \ ,
\end{split}
\ee
where the covariant derivatives act as usual, i.e. $D_+ {\cal A}^{( m'\!, n)}_{+ -}=\partial_+{\cal A}^{ (m'\!, n)}_{+ -}-[A_+, {\cal A}^{( m'\!, n)}_{+ -}]$ and $D_- {\cal A}^{( m,  n')}_{+ -}=\partial_+{\cal A}^{( m,  n')}_{+ -}-[A_-, {\cal A}^{( m,  n')}_{+ -}]$.
As a result the equations of motion can be written solely in terms of the gauge fields.

\no
In order to proceed with the calculation one should find a classical solution to \eqn{con1}.
Unlike \eqn{apam1} for the case of the single current, these are much more difficult to handle due to their nonlinearity.
However, since as before we aim at setting eventually the coupling $s$ to zero, we only need the solution valid to ${\cal O}(s)$.
Hence we find that
\be
\label{solllu}
\begin{split}
& A_+= i \big(\l^{-1}\mathbb{1} - D\big)^{-1} J_+ - n s \big(\l^{-1}\mathbb{1} - D\big)^{-1} {\cal A}^{(m,n')}_{+-}  + {\cal O}(s^2)\ ,
\\
&
A_- =- i \big(\l^{-1}\mathbb{1} - D^T\big)^{-1} J_- - ms \big(\l^{-1}\mathbb{1} - D^T\big)^{-1} {\cal A}^{(m'\!,n)}_{+-}  + {\cal O}(s^2)\ .
\end{split}
\ee
We emphasize  that in the second term in each of  the above expressions, for the gauge fields entering the definitions
\eqn{defin1} we should use the leading order expressions given by the first leading terms. The reason is that
these terms are already multiplied by $s$ and we only keep terms up to linear order in that parameter.
Substitution into the action \eqn{action-gen} we obtain that
\be
\label{djkg11ge}
S= S_k(g) + {k\ov \pi} \int d^2\s\  J_+ (\l^{-1}\mathbb{1} - D^T)^{-1} J_-
- {k s\ov \pi} \int d^2\s\ {\cal A}_{+-}^{(m,n)} + {\cal O}(s^2)\  ,
\ee
where as before in ${\cal A}_{+-}^{(m,n)}$ the leading order expressions for the gauge fields in \eqn{solllu} should be used.
Also, the first two terms are nothing  but the $\l$-deformed $\s$-model action as in \eqn{djkg11}.
The above expression gives the form of the $\l$-dressed operator ${\cal O}^{(m,n)}$ in \eqn{form-op}. It is simply
given by
\be
\label{amnpmdre}
{\cal O}^{(m,n)}_{\l} = {\cal A}^{(m,n)}_{+ -}
=S_{a_1\dots a_m ;b_1\dots b_n}A_+^{a_1}\dots A_+^{a_m}  A_- ^{b_1}\dots  A_-^{b_n}
\ ,
\ee
where we have used the definition \eqn{amnpm}.
 Obviously, for small values of $\l$
the  ${\cal O}_\l^{(m,n)}$ reduces to the operator  ${\cal O}^{(m,n)}$ in \eqn{form-op} up to a $\l$-dependent constant that does not affect the operator's anomalous dimension.

\no
We note in passing that (\ref{djkg11ge}) remains invariant under the generalized symmetry $k\to-k$, $\l\to\l^{-1}$, $g\to g^{-1}$, $s\to s\l^{m+n}$, or in terms of the effective coupling
 \begin{equation}
 k\to-k,\qq \l\to\l^{-1},\qq g\to g^{-1},\qq \tilde{\l}\to \tilde{\l}/\l^{m+n}\ .
  \label{symsym}
 \end{equation}
 This symmetry must be reflected to all physical quantities.
For the line element we have
\begin{equation}
ds^2=G_{\l\l}d\l^2+G_{\tilde{\l}\tilde{\l}}d\tilde{\l}^2+2G_{\l\tilde{\l}}d\l d\tilde{\l}\ , \label{linel}
\end{equation}
which up to linear order in $\tilde{\l}$ must be invariant under (\ref{symsym}). We are interested for the $\tilde{\l}=0$ limit in which the $G_{\l\tilde{\l}}$ being linear in $\tilde{\l}$, does not contribute. Furthermore, at this limit the first term transforms independently and thus is itself invariant under the transformation (\ref{symsym}).
Invariance of $G_{\tilde{\l}\tilde{\l}}$ under (\ref{symsym}) gives the condition $G_{\tilde{\l}\tilde{\l}}(\l)=\l^{-2m-2n}G_{\tilde{\l}\tilde{\l}}(1/\l)$ up to an overall sign. This is indeed satisfied by our metric component $G_{\tilde{\l}\tilde{\l}}$ in \eqn{memtrr}.

\subsection{The RG flow equations}

As before we choose the group element $g = e^{\s^+ \th_+ + \s^- \th_-}$ for two elements in the Cartan subalgebra of $G$,
so that again $J_{\pm}=-i \th_\pm$  and $D=\mathbb{1}$.
Furthermore, the expressions  for the gauge fields on the solution take the form
\be
\begin{split}
\label{sol-1}
& A_+^{(0)}= {\l \ov 1-\l}\big(\th_+ -n\,s\,{\cal A}^{(0)( m,  n')}_{+ -}\big) +{\cal O}(s^2)\ ,
\\
&A_-^{(0)}= -{\l \ov 1-\l}\big(\th_- +m\,s\,{\cal A}^{(0)(  m'\!,  n)}_{+ -}\big)+{\cal O}(s^2) \ .
\end{split}
\ee
The notation ${\cal A}^{(0)(  m'\!,  n)}_{+ -}$ and ${\cal A}^{(0)(  m\!,  n')}_{+ -}$ should be self-explanatory,
namely in  the definition \eqn{defin1} we should put for $A_{\pm}$ their classical
values, in particular the leading $s$-independent term of \eqn{sol-1}.

\no
Also note that \eqn{sol-1} should satisfy \eqn{eomAinitial11} as well. This is
warranted if ${\cal A}^{(0)( m',  n)}_{+ -}$ and ${\cal A}^{(0)( m,  n')}_{+ -}$ belong to the
Cartan subalgebra as well, that is similarly to the $\th_\pm$'s. This is indeed the case since the
tensor components $S_{a a_1\dots a_{m-1};b_1\dots b_n}$ and
$S_{a_1\dots a_m;bb_1\dots b_{n-1}}$ vanish if $a_i$ and $b_j$ are
Cartan indices, unless $a$ and $b$ are so as well.

We are now in the position to write down the expression for the action \eqn{djkg11ge} on the classical solution \eqn{sol-1}.
To linear order in $s$ we obtain that
\be
\label{class-Lag}
{\cal L}^{(0)}=-{k \ov 2 \pi}\left({1+\l \ov 1-\l}\th _+ \th _-+2 s\,(-1)^n \Big({\l \ov 1-\l}\Big)^{m+n}\th^{(  m,  n)}_{+ -}\right)+{\cal O}(s^2)\ ,
\ee
where we have used the definition
\be
\label{thmnpm}
\th^{(m,n)}_{+ -}=S_{a_1\dots a_m ;b_1\dots b_n}\th_+^{a_1}\dots\th_+^{a_m}  \th_-^{b_1}\dots \th_-^{b_n} \ ,
\ee
which is analogous to that in \eqn{amnpm}. Note that \eqn{class-Lag} reduces to \eqn{gg233} for $m=1,n=0$.

\no
Next we calculate the fluctuations of \eqn{eomAinitial11} around the classical solution keeping only
terms linear in them.
The result for the first equation of \eqn{eomAinitial11} reads
\ba
\label{fluct-1}
&& \Big( - \l \d_{ab}\del_- -(\tilde A_-^{(0)})_{ab}+m (m-1)s\l   \big({\cal A}^{(0)(  m''\!,  n)}_{+ -}\big)_{ab } \del_+
-im s \l  f_{abc}   \big({\cal A}^{(0)(  m'\!,  n)}_{+ -} \big)_{c}
\nonumber\\
&&\phantom{xxxxxx} +m (m-1)s\l  (\tilde A^{(0)}_+)_{ac}  \big({\cal A}^{(0)(  m''\!,  n)}_{+ -} \big)_{cb }  \Big)\d A_+^b
\\
&&
+ \Big( \d_{ab}\del_+ +(\tilde A^{(0)}_+)_{ab} + mns\l  \big({\cal A}^{(0)(  m'\!,  n')}_{+ -} \big)_{ab}\del_+
+mns \l (\tilde A^{(0)}_+)_{ac}  \big({\cal A}^{(0)(  m'\!,  n')}_{+ -} \big)_{cb} \Big)\d A_-^b=0\ .
\nonumber
\ea
In a similar fashion, the fluctuation of the second equation in \eqn{eomAinitial11} gives
\ba
\label{fluct-2}
&& \Big( - \l \d_{ab} \del_+ -(\tilde A^{(0)}_+)_{ab}+n (n-1)s\l   \big({\cal A}^{(0)(  m\!,  n'')}_{+ -}\big)_{ab }
\del_- -in s \l  f_{abc}   \big({\cal A}^{(0)(  m\!,  n')}_{+ -}\big)_{c}
\nonumber\\
&&\phantom{xxxxxx} +n (n-1)s\l  (\tilde A^{(0)}_-)_{ac} \big({\cal A}^{(0)(  m\!,  n'')}_{+ -}\big)_{cb }  \Big)\d A_-^b
\\
&&
+ \Big( \d_{ab}\del_- +(\tilde A^{(0)}_-)_{ab}+mns\l \big({\cal A}^{(0)(  m'\!,  n')}_{+ -}\big)_{ba}\del_-
+mns \l (\tilde A^{(0)}_-)_{ac} \big({\cal A}^{(0)(  m'\!,  n')}_{+ -}\big)_{bc} \Big)\d A_+^b=0\ .
\nonumber
\ea
We have also defined the following double-primed quantities
\be
\label{defin2}
\begin{split}
&\big({\cal A}^{( m''\!, n)}_{+-}\big)_{ab}=
S_{a b a_1\dots a_{m-2} ;b_1\dots b_n} A_+^{a_1}\dots A_+^{a_{m-2}} A_- ^{b_1}\dots  A_-^{b_n}\ ,
\\
& \big({\cal A}^{( m, n'')}_{+-}\big)_{ab}=
S_{a_1\dots a_{m} ;ab b_1\dots b_{n-2}} A_+^{a_1}\dots A_+^{a_m} A_- ^{b_1}\dots  A_-^{b_{n-2}}\ ,
\\
&\big({\cal A}^{( m',  n')}_{+-}\big)_{ab}=S_{a a_1\dots a_{m-1} ;b b1\dots b_{n-1}}A_+^{a_1}\dots A_+^{a_{m-1}}
A_- ^{b_1}\dots  A_-^{b_{n-1}}\ ,
\end{split}
\ee
where as before a prime or two imply that  two  indices in the tensor $S$ are not contracted.
The fluctuations equations \eqn{fluct-1} and \eqn{fluct-2} can be rewritten in the form \eqn{dhatt} with $\hat D = \hat C + \hat F$.
In momentum space we have
\be
\hat C=\hat C_0+s\, \hat C_1 \ ,\qq  \hat F=\hat F_0+s \, \hat F_1\ ,
\ee
where
\be\label{C0F0}
\hat C_0 = \left(
           \begin{array}{cc}
              -\l p_+ &  p_-  \\
              p_+ &  -\l p_-  \\
           \end{array}
         \right)\ , \qq
\hat F_0 =\left(
           \begin{array}{cc}
          - \tilde A_+^{(0)} &  \tilde A_-^{(0)} \\
        \tilde A_+^{(0)} & -\tilde A_-^{(0)}  \\
           \end{array}
         \right)\  .
\ee
and
\be\label{C1F1}
\hat C_1 = \left(
           \begin{array}{cc}
              -\l \,E\, p_- &  \l\, B \,p_-  \\
              \l \,\tilde B \,p_+ &  -\l \,\tilde E \,p_+  \\
           \end{array}
         \right)\ , \qq
\hat F_1 =\left(
           \begin{array}{cc}
          -\l\, F&   \l \,C  \\
       \l\, \tilde C& -\l\, \tilde F  \\
           \end{array}
         \right)\  .
\ee
Each of the entries in the matrices of \eqn{C0F0} and \eqn{C1F1} is itself a matrix having two indices. Namely,
the matrix components are
\be
\begin{split}
\label{def-def}
&B_{ab}=m n \big( {\cal A}^{(0)( m'\!,  n')}_{+ -}\big)_{ba} \ ,\qquad
E_{ab}=- n (n-1) \big({\cal A}^{(0)(  m\!,  n'')}_{+ -}\big)_{ab}\ ,
\\
&
\tilde B_{ab}=m n  \big({\cal A}^{(0)(  m'\!,  n')}_{+ -}\big)_{ab}\  ,  \qq
\tilde E_{ab}=- m (m-1) \big({\cal A}^{(0)(  m''\!,  n)}_{+ -}\big)_{ab}\ ,
\\
&F_{ab}=-  n (n-1) (\tilde A_-^{(0)})_{ac} \big({\cal A}^{(0)(m\!,  n'')}_{+ -}\big)_{cb}
+ i n f_{abc} \big({\cal A}^{(0)(  m\!,  n')}_{+ -}\big)_c\ ,
\\
&\tilde F_{ab}= - m (m-1) (\tilde A_+^{(0)})_{ac} \big({\cal A}^{(0)(  m''\!,  n)}_{+ -}\big)_{cb}
+i m f_{abc} \big({\cal A}^{(0)( m'\!,  n)}_{+ -}\big)_{c} \ ,
\\
&C_{ab}=mn (\tilde A^{(0)}_-)_{ac} \big({\cal A}^{(0)(  m'\!,  n')}_{+ -}\big)_{bc} \ ,\qq
\tilde C_{ab}=mn (\tilde A^{(0)}_+)_{ac} \big({\cal A}^{(0)(  m'\!,  n')}_{+ -}\big)_{cb}\ .
\end{split}
\ee
Given these expressions one can straightforwardly calculate the trace of the matrix $(\hat C^{-1} \hat F)^2$ in \eqn{fhhreff} keeping only the terms that will give rise to non-vanishing contributions upon the angular integration.
The latter will contribute an extra factor of $2\pi$.
In this way we obtain that
\be
\begin{split}
\label{trsquared}
& \Tr{(\hat C^{-1} \hat F)^2}= \Tr(\hat C^{-1}_0 \hat F_0)^2
\\
&\phantom{xxxxx}
+2s\Big( \Tr{(\hat C^{-1}_0 \hat F_0\hat C^{-1}_0\hat F_1)}-\Tr{\big((\hat C^{-1}_0 \hat F_0)^2\hat C^{-1}_0\hat C_1\big)}\Big) + {\cal O}(s^2)\ .
\end{split}
\ee
Evaluating each of the traces in the right hand side of \eqn{trsquared} separately one gets  that
\be
\begin{split}
\label{1st}
\Tr{(\hat C^{-1}_0 \hat F_0)^2}=-{8 c_G \ov r^2} {\l^2\ov (1-\l^2)^2}\th_+\th_- - {8(m+n)c_G\l s\ov r^2(1-\l)(1+\l)^2} {\cal A}^{(0)(  m\!,  n)}_{+ -}\ ,
\end{split}
\ee
and
\be
\begin{split}
\label{2nd}
&\Tr(\hat C^{-1}_0 \hat F_0\hat C^{-1}_0\hat F_1)={4 \l \ov r^2 (1-\l)(1+\l)^2}
\Big(\Tr\big(\tilde A_-^{(0)} F+  \tilde A_+^{(0)} \tilde F\big)
\\
&\qq\qq\qq\qq -\l \Tr(\tilde A_+^{(0)}  C + \tilde A_-^{(0)} \tilde C ) \Big)\, .
\end{split}
\ee
For the last trace one obtains
\be
\label{3rd}
\begin{split}
& \Tr((\hat C^{-1}_0 \hat F_0)^2\hat C^{-1}_0\hat C_1)={4 \l \ov r^2 (1-\l)(1+\l)^3}
\Big( \Tr \big( \tilde B  \tilde A_-^{(0)} \tilde A_+^{(0)}
+ \tilde E  \tilde A_+^{(0)} \tilde A_+^{(0)}
\\
& \phantom{xxxxx} +E  \tilde A_-^{(0)} \tilde A_-^{(0)} +  B  \tilde A_+^{(0)} \tilde A_-^{(0)})  -\l \Tr( B  \tilde A_-^{(0)} \tilde A_+^{(0)}
+\tilde B  \tilde A_+^{(0)} \tilde A_-^{(0)} ) \Big) \ .
\end{split}
\ee
As previously, the various traces appearing in \eqn{2nd} and \eqn{3rd} should be evaluated case by case since their result depends on the particular form of the operator chosen and in particular on the choice for the tensor $S$.

 \no
It can be easily see that  in the last term of \eqn{djkg11ge} we have that
 ${\cal A}^{(m,n)}_{+-}\sim {\cal O}^{(m,n)}$ for small $s$.
Hence, the $\s$-model action \eqn{djkg11ge} for small $\l$ and $s$ becomes
\be
\label{djkg11g}
\begin{split}
&
S= S_k(g) + {k\ov \pi} \int d^2\s \Big(\l J^a_+ J^a_- + \tilde \l {\cal O}^{(m,n)} \Big) + \cdots \ ,
\end{split}
\ee
where we have introduced the effective coupling
\be
\label{redfned}
 \tilde \l \sim s \l^{m+n} \ .
 \ee
This is the analog of \eqn{djkg11p} for the single current case.
Hence, by taking the limit $\tilde \l\to 0 $ we will find the anomalous
dimension of  ${\cal O}_\l^{(m,n)}$.

This will be done by employing \eqn{singleJ} where $\b^{\tilde \l}$ now should correspond
to this operator and the metric component should be
\be
\label{metricll}
 g^{(0)}_{\tilde\l \tilde\l}\sim {1\ov (1-\l^2)^{m+n}}\ .
 \ee
The overall coefficient is irrelevant, but nevertheless it can be found in appendix \ref{zamome}, where this metric has been
 computed. It remains to compute $\b^{\tilde \l}$. However, this seems hard in general since for an arbitrary tensor $S$ in the operator \eqn{form-op} we expect a mixing of operators under the RG flow
 even if this tensor corresponds to an irreducible representations of $G$.

 \no
 In the following section, we concentrate on important cases where such an operator mixing does not occur and we compute their corresponding  anomalous dimensions.

\no
Before we proceed we address two issues.
First we note that the background field method in the presence of the, generically irrelevant, operators ${\cal O}^{(m,n)}$,
may have subtleties. However, in our
case we are not interested in obtaining information for the running of $\tilde \l$ to all orders in $\tilde \l$, but only to a linear one for small values of it. In addition, it should be mathematically consistent to set $\tilde \l=0$.
We will see in the  examples below involving irrelevant operators that it is  indeed consistent to set  $\tilde \l=0$, as in the case for the single current.

\no
The second issue concerns the very form of the ansatz in \eqn{anssww}. This clearly is not the most general form of
a gauge invariant operator one may consider to add. One may add terms with multiple covariant derivatives acting on $\tilde g$
and/or the gauge field strength $F_{+-}$. Upon gauge fixing $\tilde g=\mathbb{1}$ such terms will give rise to descendant-like operators with  terms
having derivatives on the gauge fields,  i.e. $\del_+A_\pm$  and the commutator between gauge fields $[A_+,A_-]$.
For small values of $\l$ these will correspond to derivatives on the currents  i.e. $\del_+J_\pm$  and the commutator between currents $[J_+,J_-]$. The resulting operators with $m(n)$ in total chiral (anti-chiral) currents and derivatives $\del_+ (\del_-)$ could in principle mix with the operator in \eqn{form-op}.
Using conformal perturbation theory we have checked that such a mixing does not occur to ${\cal O}(1/k)$ but it may do so at higher orders for which we are not interested in the present paper. The reason is the following:
Quite generally,
consider schematically the overlaps $\langle {\cal O}_i  {\cal O}_j\rangle$, where  ${\cal O}_i\in \{
{\cal O}^{(m,n)},  \tilde {\cal O}^{(m,n)}\}$, with $\tilde {\cal O}^{(m,n)}$ being descendant-like operators.
This overlap is of the form
\be
\langle {\cal O}_i(x_1)  {\cal O}_j(x_2)\rangle = {1\ov x_{12}^{2 m} \bar x_{12}^{2 n}}\Big(A_{ij} + B_{ij} \ln {\e^2\ov |x_{12}|^2}\Big )\ ,
\ee
where the matrices  have the form
\be
A=\left(
    \begin{array}{cc}
      {\cal O}(1) &   {\cal O}(1/\sqrt{k}) \\
      {\cal O}(1/\sqrt{k}) & {\cal O}(1) \\
    \end{array}
  \right)\ ,\qq B=\left(
    \begin{array}{cc}
      {\cal O}(1/k) &   {\cal O}(1/k^{3/2}) \\
    {\cal O}(1/k^{3/2}) &    {\cal O}(1/k) \\
    \end{array}
  \right)\ ,
\ee
 as far as their order in $1/k$. The anomalous dimension matrix is given by $\g= A^{-1} B$. Clearly, up to ${\cal O}(1/k)$ there is no mixing since the anti-diagonal entries of $\g$ are of ${\cal O}(1/k^{3/2})$ and thus they can be ignored.

 \subsection{Important examples}

We are now focused on  some basic important examples.  These concerns operators which in the UV limit at $\l=0$
 are a general string of purely chiral current operators as well as mixed current operators.
 For the former class we generally conclude that the anomalous dimension is zero!
 For the latter ones there is in general mixing related to the associated representation theory operators. There is not such
 mixing for the operator having two $J_+$'s and one $J_-$. For this particular operator we find that the anomalous dimension is the same as that of  the operator $J_+ J_-$ driving the deformation from the CFT point.
 We have also checked that the anomalous dimensions of operators factorizing into a chiral and an anti-chiral part also vanishes.

 \no
 We will mainly concentrate to the case of the $SU(N)$ group for which we have collected some useful formulas in the appendix
 \ref{group}.

\subsubsection{The chiral operator  ${\cal O}^{(m,0)}$}

 The operator whose anomalous dimension we are interested in is at the CFT point of the form
 \be
 \label{om0}
 {\cal O}^{(m,0)}=d^{(m)}_{a_1\dots a_m}J_+^{a_1}\dots J_+^{a_m} \ ,
 \ee
 where $d^{(m)}_{a_1\dots a_m}$ is the completely symmetric rank-$m$ tensor of $SU(N)$.
 At the CFT point this is a primary field with dimension equal to $m$ \cite{BaisNorm} where $m\geqslant 3$.
 For $m=2$ this field is proportional to the energy momentum tensor.
 Its $\l$-dressed version will be given by \eqn{amnpmdre} with $n=0$.
For this class of operators certain simplifications occur when we apply the general formalism we have developed and
in addition no mixing with other operators will occur as we will readily see.
Indeed, most of the matrices in \eqn{def-def} vanish. Then, for the non-vanishing traces appearing in \eqn{2nd} and \eqn{3rd},
we have that
\be
\label{ident}
\begin{split}
&  \Tr(\tilde A^{(0)}_+  \tilde F) = m \big(c_{G} + (m-1) \D_m\big) {\cal A}^{(0)(m,0)}_{+-} = 0 \ ,
\\
& \Tr(\tilde E  \tilde A_+^{(0)} \tilde A_+^{(0)}) = m(m-1) \D_m  {\cal A}^{(0)(m,0)}_{+-} = - m c_G  {\cal A}^{(0)(m,0)}_{+-} \ ,
 \end{split}
 \ee
 where in the last step we have used \eqn{useidd} valid for $m=2,3,\dots$.

\no
Then \eqn{fhhreff} with \eqn{class-Lag} computed with $n=0$ is given by
\be
\label{hgg21m}
\begin{split}
 & -\cL_{\rm eff} = -{k\ov 2\pi} \left({1+\l\ov 1-\l} \th_+\th_- +2
 {\tilde \l \ov (1-\l)^m}\th^{(  m, 0)}_{+ -}\right)
 \\
& \qq\quad  -{c_G  \ov 2 \pi} {\l^2\ov (1-\l^2)^2} \ln \m^2 \left( \th_+\th_-
+{m \tilde \l \ov (1-\l)^{m-1}(1+\l)} \th^{(  m, 0)}_{+ -}\right) +{\cal O}(\tilde \l^2)\ ,
\end{split}
\ee
where the effective coupling is $\tilde \l\sim s\l^m$.
Then, demanding that $\del_{\ln \m^2} \cL_{\rm eff}=0$  we obtain to leading order in the $1/k$
the expression for $\b^\l$ in \eqn{systrg} as well as
 \be
 \b^{\tilde \l} = {c_G\ov 2k} {m \tilde \l \l^3\ov (1-\l)(1+\l)^3} + {\cal O}(\tilde \l^2)\ .
 \ee
Using \eqn{singleJ} with the metric entering given by \eqn{metricll} with $n=0$,
we find that
\be
\g_{{\cal O}_\l^{(m,0)}} =0\ .
\label{jjhjfh}
\ee
Before  commenting on this result we mention that for the case $m=1$ \eqn{useidd} does not make sense.
In that particular case we end up with the result for the anomalous dimensions of the single current in \eqn{ansingg}.

\no
The result \eqn{jjhjfh} is robust and has a simple explanation. Recall that the $\l$-deformed action has two well defined limits involving
$k\to \infty$ and $\l\to \pm 1$ in such a way that $k(1-\l)$ and $k (1+\l)^3$ remain finite.
These are the non-Abelian and pseudo-chiral limits, respectively \cite{Sfetsos:2013wia,Georgiou:2016iom}.
The above suggest that the anomalous
dimension of any operator ${\cal O}$ should have the form
\be
\g_{\cal O} = {c_G\ov k} {\l^n f(\l)\ov (1-\l)(1+\l)^3}\ ,
\ee
where $n$ is a non-negative integer whose value is dictated by the leading order perturbative in $\l$ result, or zero if the operator
itself has an $1\ov k$ expansion even at the CFT point for $\l=0$. The overall function $f(\l)$ is analytic in $\l$.
The symmetry \eqn{symmkl} should be encoded in the anomalous dimension of the operator which then should remain invariant. That gives the condition
\be\label{analytic}
\l^{2(2-n)}f(1/\l) =f(\l)\ .
\ee
For $n=0,1,2$, the function $f(\l)$ is a fourth, second and zeroth order polynomial. However, for $n\geqslant 3$ equation \eqn{analytic} can not hold , unless $f(\l)=0$ leading to a vanishing anomalous dimension.
Hence, if one finds a vanishing anomalous dimension up to ${\cal O}(\l^2)$
then this will vanish to all orders in $\l$ as well. This statement holds up to
${\cal O}(1/k)$ and similar statements can be made for higher order terms in the large $k$-expansion.
We have performed perturbative consistency check of the above in section 6 with complete agreement with the above statements.

\no
An important comment is in order. The operator $J_+^aJ_+^a$ is at the CFT limit, that is  when $\l=0$, proportional to the chiral component of the energy momentum tensor. In the $\l$-deformed theory one may readily check that the role of the energy momentum tensor is played by the deformation of $J_+^aJ_+^a$, namely
 ${\cal O}_\l^{(2,0)}$. Hence,
\be
\label{tttpp}
{\cal O}_\l^{(2,0)}={\cal A}_{+-}^{(2,0)}\sim J_+ (1-\l D^T)^{-1}(1-\l D)^{-1} J_+ \sim T_{++}\ .
\ee
The last proportionality relation to $T_{++}$ follows by simply evaluating the energy momentum tensor for
the $\l$-deformed model action \eqn{djkg11} (with $s=0$).
As in any $\s$-model this is classically chirally conserved, i.e. $\del_- T_{++}=0$. A less trivial statement is that
the following sequence of chiral conservation laws
\be
\label{hh29}
\del_- {\cal O}^{(m,0)}_\l = 0\ ,\qq m=2,3,\dots \ .
\ee
holds, in which the  chiral conservation law for $T_{++}$ is just the first member.
This is a consequence of the fact that the classical equation of motion for the $\l$-deformed model can be cast as
\be
\del_\mp A_\pm= \mp {1\ov 1+\l} [A_+,A_-]  \ ,
\ee
as well as of the group theoretical identity \eqn{useidd0}.\footnote{The fact
that $\g_ {{\cal O}^{(m,0)}}=0$ is consistent with the classical statement \eqn{hh29}.
Equation \eqn{hh29} does not hold quantum mechanically. The reason is  that the form of the operator
we have used, i.e. given  by \eqn{apam1} (with $s=0$)  and consequently by \eqn{amnpmdre}, receives $1/k$-corrections.
This is in agreement with the fact that the theory is not conformally invariant already at ${\cal O}(1/k)$.}

\no
Obviously, the anomalous dimension of the operator ${\cal O}^{(0,n)}_\l$ made up purely of anti-chiral currents vanishes as well.
In addition, even though this is less obvious, we have checked that the operator ${\cal O}^{(m,0)}_\l {\cal O}^{(0,n)}_\l$ has also vanishing anomalous dimension.
We choose not to present the details of the computation which nevertheless are similar to those presented in this subsection.

\subsubsection{The mixed operator ${\cal O}^{(2,1)}$}

 The operator whose anomalous dimension we are interested in is of the form
 \be
 {\cal O}^{(2,1)}=d_{abc}J_+^{a}J_+^b J_-^c \ ,
 \ee
 where $d_{abc}$ is the completely symmetric tensor of $SU(N)$ of rank three.
This operator cannot mix with others and its $\l$-dressed form is given by \eqn{amnpmdre} for $m=2$ and $n=1$.
Recall that the field $Q^a=d_{abc}J_+^b J_+^c$ is primary with dimension equal to $2$ \cite{BaisNorm}.
Hence,  the operator ${\cal O}^{(2,1)}$ at the CFT point is a primary field with holomorphic and anti-holomorphic dimensions (in a Euclidean regime language) dimensions equal to $2$ and $1$, respectively.

\no
Setting $n=1$ some matrices in \eqn{def-def} vanish or simplify.
Then we have for the various traces appearing in \eqn{2nd} and \eqn{3rd} that
\be
\begin{split}
& \Tr(\tilde A_-^{(0)} F)=\Tr(\tilde A_+^{(0)} \tilde F) =\Tr(\tilde B \tilde A_-^{(0)} \tilde A_+^{(0)})
= - \Tr(\tilde E \tilde A_+^{(0)} \tilde A_+^{(0)})
\\
&
 \phantom{x} = \Tr(B \tilde A_-^{(0)} \tilde A_+^{(0)}) = \Tr(\tilde B \tilde A_+^{(0)} \tilde A_-^{(0)})
=\Tr( B \tilde A_+^{(0)} \tilde A_-^{(0)})
\\
&
 \phantom{x} =\Tr(\tilde A_+^{(0)} C) = \Tr(\tilde A_-^{(0)} \tilde C)
\\
&\phantom{x} =  - c_G {\l^3\ov (1-\l)^3} \th_{+-}^{(2,1)} +  {\cal O}(s)\ ,
\end{split}
\ee
where we kept only the leading order result in $s$ since these terms are already multiplied by $s$ in \eqn{trsquared}
(via \eqn{2nd} and \eqn{3rd}).
Then \eqn{fhhreff} with \eqn{class-Lag} computed with $m=2$ and $n=1$ becomes
\be
\label{hgg21}
\begin{split}
 & -\cL_{\rm eff} = -{k\ov 2\pi} \left({1+\l\ov 1-\l} \th_+\th_- -2  {\tilde \l\ov(1-\l)^3} \th_{+-}^{(2,1)}\right)
 \\
& \qq\quad -{c_G  \ov 2 \pi}{\l\ov (1-\l^2)^2} \ln \m^2 \left(\l \th_+\th_-  - {\tilde \l(2+\l+2 \l^2)\ov
(1-\l)^2(1+\l)} \th_{+-}^{(2,1)} \right) + {\cal O}(\tilde \l^2) \ ,
\end{split}
\ee
where in this case the effective coupling is $\tilde \l= s\l^3$.
Demanding that $\del_{\ln \m^2} \cL_{\rm eff}=0$  we obtain to leading order in $1/k$
the expression for $\b^\l$ written in \eqn{systrg} and that
 \be
 \b^{\tilde \l} = -{c_G\ov 2k} {\tilde \l \l\big(2- \l(2+\l)\big)\ov (1-\l)(1+\l)^3} + {\cal O}(\hat \l^2)\ .
 \ee
Using \eqn{singleJ} with the metric entering given by \eqn{metricll} again with $m=2$ and $n=1$
we finally find that
\be
\g_ {{\cal O}_\l^{(2,1)}} = -{2 c_G\ov k} \frac{ \lambda (1-\lambda(1-\lambda))}{(1-\lambda)(1+\lambda)^3}\ ,
\label{jjhjf}
\ee
 which is the same as that in for $\g_{J_+J_-}$ in \eqn{jhjf}.

\section{$\l$-deformations with different current algebra levels}

In this section, we will use our general formalism in order to calculate the anomalous dimensions of current composite operators in models for which the levels of the chiral and anti-chiral algebras are different.
The main motivation is that such models generically have fixed points in the IR corresponding to new CFTs.
The first such model was presented in \cite{Georgiou:2017jfi} in which one starts with two WZW models at different levels $k_1$ and $k_2$ and via a gauging procedure involving two sets of gauge fields $A_\pm$ and $B_\pm$ one constructs the all-loop effective action of two mutually interacting WZW models. The terms driving the models away from the CFT point are $J_{1+}J_{2-}$ and
$J_{2+}J_{1-}$, where the index $1$ or $2$ indicates that they refer to the first or
the second WZW model and the corresponding levels are $k_1$ and $k_2$.
We may simplify further the model by consistently setting the coupling of the second of these terms to zero as we will explain below.
Then, it turns out that it is consistent to consider operators of a form
similar to \eqn{form-op} and given by
\be
\label{form-op-uneq}
{\cal O}^{(m,n)}
=S_{a_1\dots a_m;b_1 \dots b_n}J_{1+}^{a_1}\dots J_{1+}^{a_m}  J_{2-} ^{b_1}\dots  J_{2-} ^{b_n}\ .
\ee
As in the single $\l$-deformations the overall tensor coefficient should be symmetric in the first $m$ indices, as well as in the last $n$ ones, separately with no symmetry property relating  the $a_i$'s and the $b_i$'s.

\no
Our starting point will be eq. (2.6) of  \cite{Georgiou:2017jfi} but with $\l_2 \rightarrow 0$ and $\l_1$ renamed to
$\l$. It turns out that in this limit, which is consistent quantum mechanically from an RG flow point of view, the leading
order term for small remaining coupling $\l$ is $J_{1+}J_{2-}$ which, as mentioned above, is the case we want to concentrate on. Then the last term in the first line of equation (2.6) of  \cite{Georgiou:2017jfi} remains finite if we first rescale $B_\pm$ as $B_\pm \rightarrow  \sqrt{\l_2}B_\pm$ and then take $\l_2\to 0$. In this limit the effective action analog of \eqn{action-gen} becomes
\be
\begin{split}
\label{action-gen-un}
& S_{k_i,\l,s}(g_i, A_\pm) =\sum_{i=1}^2 S_{k_i}(g_i)
+{1\ov \pi} \int d^2\s \ \Tr \big(k_1 A_- \del_+ g_1 g_1^{-1}   -k_2  A_+ g_2^{-1} \del_- g_2\big) \\
&\qq\qq\qq - {\sqrt{k_1 k_2} \ov \pi}  \int d^2\s\ \left(\l^{-1} \Tr\big( A_+ A_-\big) + s  {\cal A}^{(m,n)}_{+ -}\right)\ ,
\end{split}
\ee
where as before the expression for ${\cal A}^{(m,n)}_{+ -}$ is given by \eqn{amnpm}.
In this procedure the gauge field $B_\pm$ has decoupled, which is the reason we have not included the term $\Tr(B_+B_-)$ in the above action, even though its overall coupling constant remains finite.

\no
The equations of motion for \eqn{action-gen-un} with respect to $A_-$ and $A_+$ are given by
\be
\label{con1-un}
\begin{split}
&
D_+ g_1\, g_1^{-1} =  (\l_0^{-1}\l^{-1}-1) A_+ + \l_0^{-1} n s   {\cal A}^{(m, n')}_{+ -}\ ,
\\
&
g_2^{-1} D_- g_2 =  - (\l_0 \l^{-1}-1)  A_- - \l_0m s {\cal A}^{( m'\!, n)}_{+ -},
\end{split}
\ee
where we have defined the covariant derivatives  $D_+ g_1=\del_{+}g_1-A_+ g_1$ and $D_- g_2=\del_{-}g_2+g_2 A_-$
and the vectors ${\cal A}^{(m, n')}_{+ -}$ and ${\cal A}^{(m', n)}_{+ -}$ are given by \eqn{defin1}.
Instead of the levels $k_1$ and $k_2$ we will use the parameters
\be
\l_0=\sqrt{{k_1\ov k_2}}\ ,\qq k=\sqrt{k_1 k_2}\ .
\ee

\no
Varying the action with respect to the group element $g_1$ and $g_2$ results into the following set of equations
\be
\begin{split}
\label{g-eom1}
\del_-( D_+ g_1 g_1^{-1}) - [A_-, D_+ g_1 g_1^{-1}]=F_{+-}\ ,\qquad
\del_-( D_+ g_2 g_2^{-1})=0 \ ,
\end{split}
\ee
or equivalently
\be
\begin{split}
\label{g-eom2}
\del_+(g_1^{-1} D_- g_1)=0\ ,\qquad
\del_+(g_2^{-1} D_- g_2) - [A_+,g_2^{-1} D_- g_2]=F_{+-}\ ,
\end{split}
\ee

\no
Substituting the constraints \eqn{con1-un} in \eqn{g-eom1} and \eqn{g-eom2} we obtain that
\be
\begin{split}
\label{eomAinitial11-un}
&\l_0 \l^{-1} \del_+  A_- -\del_-A_+
= \l_0 \l^{-1}[ A_+, A_-] -\l_0 m \, s \, D_+ {\cal A}^{( m'\!, n)}_{+ -} \ ,
\\
& \l_0^{-1} \l^{-1} \del_-  A_+ -\del_+A_-
=-   \l_0^{-1} \l^{-1}  [A_+,A_-] -  \l_0^{-1}n \, s \, D_- {\cal A}^{( m,  n')}_{+ -} \ ,
\end{split}
\ee
where the covariant derivatives act as explained below \eqn{eomAinitial11}.
Hence, the equations of motion have been  written in terms of the gauge fields only.

\no
The constraints \eqn{con1-un} can be easily solved to give
\be
\begin{split}
& A_+= i \l_0\l J_{1+} - n s\l {\cal A}_{+-}^{(m,n')} + {\cal O}(s^2)\ ,
\\
&
 A_-=- i \l_0^{-1}\l J_{2-} - m s\l {\cal A}_{+-}^{(m'\!,n)} + {\cal O}(s^2)\ ,
\end{split}
\ee
where in ${\cal A}_{+-}^{(m,n')} $ and in ${\cal A}_{+-}^{(m'\!,n)}$ above only the leading order first terms
should be used in their definition \eqn{defin1}. Then substitution into \eqn{action-gen-un} gives the action
 \be
\label{djkg11g-unh}
\begin{split}
&
S= \sum_{i=1}^2 S_{k_i}(g_i) + {k\ov \pi} \int d^2\s \Big(\l J^a_{1+} J^a_{2-} - s {\cal A}_{+-}^{(m,n)}
\Big)+  {\cal O}(s^2)  \ .
\end{split}
\ee
Note that since in the action above in ${\cal A}_{+-}^{(m,n)}$ only the leading order expressions in $s$ should be used, the $\l$-dressing of the gauge fields is just a trivial overall constant unlike the case for the gauge fields for the single $\l$-deformed models in \eqn{solllu}.
Consequently, the operator \eqn{form-op-uneq} does not get change upon the $\l$-deformation.

\subsection{The RG flow equations}

In order to proceed with the calculation one should find a classical solution to \eqn{eomAinitial11-un}. However, as we did in the previous cases we only need to find a solution valid to order ${\cal O}(s)$. This can
be easily obtained if we  choose the group elements $g_i = e^{\s^+ \th_+^{(i)} + \s^- \th_-^{(i)} },\, i=1,2$ with the elements $\th_\pm^{(i)} $ belonging in the Cartan subalgebra of $G$,
so that $J_{i\pm}=-i \th_\pm^{(i)}$.
Furthermore, the expressions  for the gauge fields on the solution take the form
\be
\begin{split}
\label{sol-1-un}
& A_+^{(0)}= \l_0 \l \th_+^{(1)} -n\,s\,\l {\cal A}^{(0)( m,  n')}_{+ -} +{\cal O}(s^2)\ ,
\\
&A_-^{(0)}= -\l_0^{-1}\l \th_-^{(2)}  -m\,s\,\l{\cal A}^{(0)(  m'\!,  n)}_{+ -}+{\cal O}(s^2) \ .
\end{split}
\ee
Notice that in  the definition \eqn{defin1} we should put for $A_{\pm}$ their classical
values \eqn{sol-1-un}.  Notice also that \eqn{sol-1-un} should also satisfy \eqn{eomAinitial11-un}. This is
guaranteed  if ${\cal A}^{(0)( m',  n)}_{+ -}$ and ${\cal A}^{(0)( m,  n')}_{+ -}$ belong to the
Cartan subalgebra  similarly to the the $\th_\pm$'s. As mentioned before, this is indeed the case since the
tensor components $S_{a a_1\dots a_{m-1};b_1\dots b_n}$ and
$S_{a_1\dots a_m;bb_1\dots b_{n-1}}$ vanish if $a_i$ and $b_j$ are
Cartan indices while $a$ or $b$ are not.

\no
To linear order in $s$ the action \eqn{action-gen-un} evaluated on the classical solution \eqn{sol-1-un} is
\be
\label{class-Lag-un}
\begin{split}
& {\cal L}^{(0)}=-{1 \ov 2 \pi}\Big( k_1\th _+^{(1)} \th _-^{(1)}  +k_2  \th _+^{(2)} \th _-^{(2)} +2 k \l  \th _+^{(1)} \th _-^{(2)}
\\
&\qq\quad + 2k  s\,(-1)^n\l_0^{m-n} \l^{m+n}\th^{(  m,  n)}_{+ -}\Big)+{\cal O}(s^2)\ ,
\end{split}
\ee
where we have used a definition similar to \eqn{thmnpm}, i.e.
\be
\label{thmnpm-un}
\th^{(m,n)}_{+ -}=S_{a_1\dots a_m ;b_1\dots b_n}\th_+^{(1)a_1}\dots\th_+^{(1)a_m}  \th ^{(2)b_1}\dots \th_-^{(2)b_n} \ .
\ee

\no
The linear fluctuations of \eqn{eomAinitial11-un} around the classical solution are in order.
From the first equation of \eqn{eomAinitial11-un} we obtain that
\ba
\label{fluct-1-un}
&& \Big( - \l_0^{-1}\l \d_{ab}\del_- - (\tilde A_-^{(0)})_{ab}+m (m-1)s \l   \big({\cal A}^{(0)(  m''\!,  n)}_{+ -}\big)_{ab } \del_+
-im s \l  f_{abc}   \big({\cal A}^{(0)(  m'\!,  n)}_{+ -} \big)_{c}
\nonumber\\
&&\phantom{xxxxxx} +m (m-1)s\l  (\tilde A^{(0)}_+)_{ac}  \big({\cal A}^{(0)(  m''\!,  n)}_{+ -} \big)_{cb }  \Big)\d A_+^b
\\
&&
+ \Big( \d_{ab} \del_+ +(\tilde A^{(0)}_+)_{ab} + mns\l  \big({\cal A}^{(0)(  m'\!,  n')}_{+ -} \big)_{ab}\del_+
+mns \l (\tilde A^{(0)}_+)_{ac}  \big({\cal A}^{(0)(  m'\!,  n')}_{+ -} \big)_{cb} \Big)\d A_-^b=0\ .
\nonumber
\ea
In a similar fashion, the  second equation in \eqn{eomAinitial11-un} gives
\ba
\label{fluct-2-un}
&& \Big( -\l_0 \l \d_{ab}\del_+ -(\tilde A^{(0)}_+)_{ab}+n (n-1)s \l   \big({\cal A}^{(0)(  m\!,  n'')}_{+ -}\big)_{ab }
\del_- -in s \l  f_{abc}   \big({\cal A}^{(0)(  m\!,  n')}_{+ -}\big)_{c}
\nonumber\\
&&\phantom{xxxxxx} +n (n-1)s \l  (\tilde A^{(0)}_-)_{ac} \big({\cal A}^{(0)(  m\!,  n'')}_{+ -}\big)_{cb }  \Big)\d A_-^b
\\
&&
+ \Big(\d_{ab}\del_- +(\tilde A^{(0)}_-)_{ab}+mns\l \big({\cal A}^{(0)(  m'\!,  n')}_{+ -}\big)_{ ba}\del_-
+mns  \l (\tilde A^{(0)}_-)_{ac} \big({\cal A}^{(0)(  m'\!,  n')}_{+ -}\big)_{bc} \Big)\d A_+^b=0\ ,
\nonumber
\ea
where the  quantities with two primes are defined in \eqn{defin2}.

\no
These fluctuations can be rewritten in the form \eqn{dhatt} with $\hat D = \hat C + \hat F$.
In momentum space we have that
\be
\hat C=\hat C_0+s\, \hat C_1 \ ,\qq  \hat F=\hat F_0+s \, \hat F_1\ ,
\ee
where
\be\label{C0F0-un}
\hat C_0 = \left(
           \begin{array}{cc}
              -\l \l_0 p_+ &  p_-  \\
              p_+ &  -\l \l_0^{-1}p_-  \\
           \end{array}
         \right)\ , \qq
\hat F_0 =\left(
           \begin{array}{cc}
          - \tilde A_+^{(0)} &  \tilde A_-^{(0)} \\
        \tilde A_+^{(0)} & -\tilde A_-^{(0)}  \\
           \end{array}
         \right)\  .
\ee
and
\be\label{C1F1-un}
\hat C_1 = \left(
           \begin{array}{cc}
              -\l \,E\, p_- &  \l\, B \,p_-  \\
              \l \,\tilde B \,p_+ &  -\l \,\tilde E \,p_+  \\
           \end{array}
         \right)\ , \qq
\hat F_1 =\left(
           \begin{array}{cc}
          -\l\, F&   \l \,C  \\
       \l\, \tilde C& -\l\, \tilde F  \\
           \end{array}
         \right)\  ,
\ee
where all matrices appearing in \eqn{C0F0-un} and \eqn{C1F1-un} are defined as in \eqn{def-def}.

One can straightforwardly calculate the $\Tr(\hat C^{-1} \hat F)^2$ in \eqn{fhhreff}.
We just need to keep only the terms giving rise to non-vanishing contributions upon the angular integration
which will contribute an extra factor of $2\pi$.
Evaluating each of the traces in the right hand side of \eqn{trsquared} separately one gets for the case at hand that
\be
\begin{split}
\label{1st-un}
\Tr{(\hat C^{-1}_0 \hat F_0)^2}=-{8 c_G \ov r^2} {\l (\l-\l_0)(\l-\l_0^{-1})\ov (1-\l^2)^2}
\left(\l \th_+^{(1)}\th_-^{(2)} + (m+n)s  {\cal A}^{(0)(  m\!,  n)}_{+ -}\right)
\end{split}
\ee
and
\be
\begin{split}
\label{2nd-un}
&\Tr (\hat C^{-1}_0 \hat F_0\hat C^{-1}_0\hat F_1)={4 \l \ov r^2 (1-\l^2)^2}
\Big((1-\l_0^{-1}\l)\Tr(\tilde A_-^{(0)} F)+ (1-\l_0\l) \Tr(\tilde A_+^{(0)} \tilde F)
\\
&\qq\qq\qq\qq +\l (\l-\l_0^{-1})\Tr(\tilde A_+^{(0)}  C) +\l (\l-\l_0) \Tr(\tilde A_-^{(0)} \tilde C)   \Big)
\end{split}
\ee
and that
\be
\label{3rd-un}
\begin{split}
& \Tr((\hat C^{-1}_0 \hat F_0)^2\hat C^{-1}_0\hat C_1)={4 \l \ov r^2 (1-\l^2)^3}
\Big( (\l-\l_0)(\l-\l_0^{-1})\Tr \big( \tilde B  \tilde A_-^{(0)} \tilde A_+^{(0)}
\\
& \qq\qq+ \tilde E  \tilde A_+^{(0)} \tilde A_+^{(0)}
 +E  \tilde A_-^{(0)} \tilde A_-^{(0)} +  B  \tilde A_+^{(0)} \tilde A_-^{(0)}\big)
 \\
 &\qq\qq -\l \l_0 (\l-\l_0^{-1})^2\Tr( B  \tilde A_-^{(0)} \tilde A_+^{(0)})
-\l \l_0^{-1} (\l-\l_0)^2\Tr(\tilde B  \tilde A_+^{(0)} \tilde A_-^{(0)} ) \Big) \ .
\end{split}
\ee
The various traces appearing in \eqn{2nd} and \eqn{3rd} should be evaluated case by case since their result depends on the particular form of the operator chosen.

For small values of the parameter $s$ the $\s$-model action \eqn{action-gen-un}  becomes
\be
\label{djkg11g-un}
\begin{split}
&
S= \sum_{i=1}^2 S_{k_i}(g_i) + {k\ov \pi} \int d^2\s \Big(\l J^a_{1+} J^a_{2-}+
+ \tilde\l {\cal O}^{(m,n)}\Big) + \cdots \ .
\end{split}
\ee
where the operator added is defined in \eqn{form-op-uneq} and the effective coupling as in \eqn{redfned}.

 Hence, by taking the limit $\tilde \l\to 0$ we will find the anomalous
dimension of  ${\cal O}^{(m,n)}$. This will be done by employing \eqn{singleJ} where $\b^{\tilde \l}$ now should correspond to this operator. The metric component is still given by \eqn{metricll}
It remains to compute $\b^{\tilde \l}$ and evaluate the anomalous dimensions, a task
 undertaken in the next subsection for the important cases consider in the case of the original $\l$-deformed model in section 3.

\subsection{Important examples}

In this section we will compute the anomalous dimensions of the same operators as in section 3.

 \subsubsection{Chiral operators ${\cal O}^{(m,0)}$}

Consider operators of the form
 \be
 \label{om0-un}
 {\cal O}^{(m,0)}=d^{(m)}_{a_1\dots a_m}J_{1+}^{a_1}\dots J_{1+}^{a_m} \ ,
 \ee
 which is similar to \eqn{om0} and at the CFT point are primary fields with dimension $m$ \cite{BaisNorm} for $m\geqslant 3$ and proportional to the energy momentum tensor for $m=2$.
Certain simplifications occur since for $n=0$ most of the matrices in \eqn{def-def} vanish.
For the seemingly non-vanishing traces appearing in \eqn{2nd-un} and \eqn{3rd-un}, we have the same relations as in \eqn{ident}.

\no
Then \eqn{fhhreff} summed with \eqn{class-Lag-un} computed at $n=0$ is given by
\be
\label{hgg21-un}
\begin{split}
 & -\cL_{\rm eff} = -{k_1 \ov 2 \pi} \th _+^{(1)} \th _-^{(1)}  -{k_2 \ov 2 \pi} \th _+^{(2)} \th _-^{(2)} - {k \ov  \pi}\l  \th _+^{(1)} \th _-^{(2)}
-{k \ov \pi} \tilde \l \l_0^{m} \th^{(  m,  0)}_{+ -}
 \\
&  \quad\quad  -{\ln \m^2  \ov 2 \pi} {\l^2\ov (1-\l^2)^2} \Bigg( c_G (\l-\l_0)(\l-\l_0^{-1})  \th _+^{(1)} \th _-^{(2)}
\\
&   \quad \quad +  m \tilde \l \l_0^{m} { \l-\l_0^{-1}\ov 1-\l^2}\Big(  c_G(1-\l^2)+\D_m (m-1)(1-\l_0 \l) \Big)  \th^{(  m, 0)}_{+ -}
\Bigg) +{\cal O}(\tilde \l^2)\ ,
\end{split}
\ee
where as before $\tilde \l= s\l^m$ is the effective coupling.
Demanding that $\del_{\ln \m^2} \cL_{\rm eff}=0$  we obtain to leading order in the $1/k$
the expression for $\b^\l$ in the case of unequal levels \cite{Georgiou:2017jfi}
\be\label{fluc-un}
\b^\l=-{c_G \ov 2k}{\l^2(\l-\l_0)(\l-\l_0^{-1})\ov (1-\l^2)^2}\ ,
\ee
 as well as the $\b$-function for the coupling $\tilde \l$
 \be
 \b^{\tilde \l} =  -{m  \l^2 (  \l-\l_0^{-1}  ) \Big(c_G (1 -  \l^2) + ( m-1) \D_m (1 -  \l  \l_0) \Big)\tilde \l\ov
 2 k (1 -  \l^2)^3  }+ {\cal O}(\tilde \l^2)\ .
 \ee
Using \eqn{singleJ} with the metric entering given by \eqn{metricll} again with $n=0$
we find that
\be
\g_ {{\cal O}^{(m,0)}} =\Big(c_G+( m-1) \D_m \Big){m\l^2 (1-\l \l_0)^2\ov k_1 (1 - \l^2)^3}\ .
\label{jjhjfh-un1}
\ee
We immediate see that for $m\geqslant 2$ the anomalous dimension of the chiral operators vanish due to the group theory identity \eqn{useidd}, that is
\be
\g_{{\cal O}^{(m,0)}} =0\ , \qq m=2,3,\dots \ .
\label{jjhjfh-un}
\ee
as in the case of equal levels in \eqn{jjhjfh-un}.

\no
However, for the anomalous dimension of a single chiral current, i.e. when $m=1$, this identity does not hold and \eqn{jjhjfh-un1} gives that
\be
\g_{{\cal O}^{(1,0)}}={c_G\l^2(\l-\l_0^{-1})^2 \ov k_2 (1 - \l^2)^3}\ ,
\ee
which is in perfect agreement  with the expression of the chiral current calculated in equation (2.9) of \cite{Georgiou:2016zyo}.

\no
Note that, since the equations of motion can be cast in the form
\be
\del_\mp A_\pm = \mp {1-\l_0^{\pm 1}\l\ov 1-\l^2} [A_+,A_-]\ ,
\ee
the classical conservation law \eqn{hh29}.

Finally, by following  the same steps, one can show that the composite operators made from an arbitrary number of anti-chiral currents $J_{2-}$ have also vanishing
anomalous dimensions. This is also, rather trivially,  the case for operators built from an arbitrary number of the currents $J_{2+}$ or $J_{1-}$ since these two currents are not present
in the operator that deforms the CFT and which is $J_{1+}$$J_{2-}$.  

\subsubsection{The mixed operator ${\cal O}^{(2,1)}$}

Consider next mixed operators
 of the form
 \be
 {\cal O}^{(2,1)}=d_{abc}J_{1+}^{a}J_{1+}^b J_{2-}^c \ ,
 \ee
 where $d_{abc}$ is the completely symmetric tensor of $SU(N)$ of rank three.
Then for the various traces appearing in \eqn{2nd} and \eqn{3rd} we have that
\be
\begin{split}
& \Tr(\tilde A_-^{(0)} F)=\Tr(\tilde A_+^{(0)} \tilde F) =\Tr(\tilde B \tilde A_-^{(0)} \tilde A_+^{(0)})
= - \Tr(\tilde E \tilde A_+^{(0)} \tilde A_+^{(0)})
\\
&
 \phantom{x} = \Tr(B \tilde A_-^{(0)} \tilde A_+^{(0)}) = \Tr(\tilde B \tilde A_+^{(0)} \tilde A_-^{(0)})
=\Tr( B \tilde A_+^{(0)} \tilde A_-^{(0)})
\\
&
 \phantom{x} =\Tr(\tilde A_+^{(0)} C) = \Tr(\tilde A_-^{(0)} \tilde C)
\\
&\phantom{x} =  - c_G\l_0 \l^3 \th_{+-}^{(2,1)} +  {\cal O}(s)\ ,
\end{split}
\ee
where we kept only the leading order result  in $s$ since these terms are already multiplied by $s$ in \eqn{trsquared}.
Then \eqn{fhhreff} with \eqn{class-Lag-un} computed with $m=2$ and $n=1$ give the effective action
\be
\label{hgg21-21}
\begin{split}
 & -\cL_{\rm eff} = -{1\ov 2\pi}\left(k_1  \th _+^{(1)} \th _-^{(1)}  + k_2 \th _+^{(2)} \th _-^{(2)} + 2  k \l  \th _+^{(1)} \th _-^{(2)} - 2 k  \tilde \l\l_0 \th^{(  2,  1)}_{+ -}\right)
 \\
& \qq\quad  -{\ln \m^2  \ov 2 \pi}{c_G\l\ov (1-\l^2)^2}
\Bigg(\l (\l-\l_0)(\l-\l_0^{-1})  \th _+^{(1)} \th _-^{(2)}
\\
&\qq\quad - \tilde \l {2 \l_0-3 \l -\l_0\l(3\l_0-5 \l+\l^3)\ov 1-\l^2}
   \th^{(  2, 1)}_{+ -}\Bigg) +{\cal O}(\tilde \l^2)\ ,
\end{split}
\ee
where the effective coupling is $\tilde \l = s\l^3$.
Demanding that $\del_{\ln \m^2} \cL_{\rm eff}=0$  we obtain to leading order in the $1/k$-expansion
the expression for $\b^\l$ in \eqn{fluc-un} and that
 \be
 \b^{\tilde \l} = - {c_G\ov  2 k}{ \tilde \l \l \big(2  -  3 \l \l_0^{-1}+  \l (5 \l - \l^3 - 3 \l_0 )\big) \ov
 (1 - \l^2)^3  }+ {\cal O}(\hat \l^2)\ .
  \ee
Using \eqn{singleJ} with the metric entering given by \eqn{metricll} again with $m=2$ and $n=1$
we find that
\be
\g_ {{\cal O}^{(2,1)}} =  c_G\lambda  \frac{  3(\l_0+\l_0^{-1})\l (1+\l^2)-2(1+4 \l^2+\l^4)}{k(1-\lambda^2)^3}\ ,
\label{jjhjf-un}
\ee
 which is the same as the anomalous dimension of the operator $J_{1+}^a J_{2-}^a$.
 The latter can be found in equation (2.16) of \cite{Georgiou:2016zyo}.


\section{ $\l$-deformations of the self- and mutual-type}

In this section we consider the $\l$-deformed model constructed in \cite{Georgiou:2018hpd}
describing simultaneous interactions of two WZW models of the self- and mutually-interacting type. At the linearized level the action is
\begin{equation}
\begin{split}
&
S_{k_1,k_2,\l,\tilde{\l}}(g_1,g_2)=S_{k_1}(g_1)+S_{k_2}(g_2)
\\
&\qq\quad+\frac{k_1}{\pi}\l\int d^2\s J_{1+}J_{1-}+\frac{k_2}{\pi}\tilde{\l}\int d^2\s J_{2+}J_{1-} + {\cal O}(\l\tilde \l)\ .
 \label{2couplingsaction}
\end{split}
\end{equation}
Various aspects of this model, along with the construction of its all-order in the parameters effective action, can be found
in \cite{Georgiou:2018hpd} where it was also shown that the RG flow equations of this model have a rich structure.

\no
We will compute the Zamolodchikov's metric for this theory, along with the anomalous dimensions of the composite operators $J_{1+}J_{2-}$ and $J_{2+}J_{1-}$ that drive the perturbation away from the CFT point. Then, by taking appropriate limits we will
find the anomalous dimensions of currents of the single deformed modes with equal or even
unequal levels which have been computed before. Complete agreement will be found.
That gives extra confidence for the validity of the procedure we used in sections 2,3 and 4.

\no
The $\b-$functions for this model can be found in eqs. (4.19) and (4.20) of \cite{Georgiou:2018hpd} and read
\be
\label{systrg1}
\begin{split}
	& \b^\l(\l,\tilde \l) = -{c_G \l(1-\l) \ov 2 \D^2}\left( k_1 \l (1-\l) -k_2 \tilde \l^2 (1+\l-\tilde \l)\right)\ ,
	\\
	&
	\b^{\tilde \l}(\l,\tilde \l) =- {c_G\tilde \l (1-\tilde\l) \ov 2 \D^2}
	\left(k_1(1-\l)\left(\tilde \l-\l(\l-\tilde\l)\right)-k_2\tilde \l^2 \right)\ ,
\end{split}
\ee
where
\be
\label{gghh}
 \D= k_1(1-\l^2) -k_2\tilde \l^2\ . 
\ee
As argued in the end of section 4.1 of \cite{Georgiou:2018hpd} it is convenient to
rewrite (\ref{2couplingsaction}) after a rescaling so that one may use available results in the
literature.
Indeed, after the rescaling $J_{i\pm}\to J_{i\pm}/\sqrt{k_i}$, $i=1,2$, then (\ref{2couplingsaction}) can be rewritten as
\begin{equation}
\begin{split}
& S_{k_1,k_2,\L}=S_{k_1}(g_1)+S_{k_2}(g_2)+\frac{1}{\pi}\int d^2\s \mathcal{J}_{+A}\L_{AB}\mathcal{J}_{-B}\ ,\quad \L=
\left(       \begin{array}{cc}
\l \mathbb{1}& 0   \\
\l_0^{-1}\tilde \l \mathbb{1} &  0\\
\end{array}
\right)\ ,
\\
& \qq \mathcal{J}^A_{\pm}=\left(J^a_{1\pm},J^{a'}_{2\pm}\right)\ , \qq
 \l_0=\sqrt{\frac{k_1}{k_2}}\ ,
  \label{lambdamatrix}
  \end{split}
\end{equation}
where both group indices $a,a'=1,2,\dots , \dim G$.
Notice here that the coupling matrix $\L$ is non-invertible. However this does not affect our calculations since no inversion operation is needed.

\subsection{The Zamolochikov metric}

We will compute the Zamolochikov metric for \eqn{2couplingsaction} for finite values of both couplings.
The general form of the Zamolochikov metric was computed in \cite{Sagkrioti:2018abh}.
 Recalling the relevant expressions and using a double index notation, we have that
\be
\label{metgggh}
ds^2 = G_{AB|CD} d\L_{AB} d\L_{CD}\ ,\quad G_{AB|CD} = \frac{\text{dimG}}{2} (\tilde g^{-1})_{AC} (g^{-1})_{BD}\ ,
\ee
where
\be
g_{AB}=(\mathbb{1}-\L^T\L)_{AB}  \ ,\qq \tilde g_{AB}=(\mathbb{1}-\L\L^T)_{AB}\ .
\label{g,tildeg}
\ee
Using the matrix $\l$ in (\ref{lambdamatrix}) we find that
\be
\begin{split}
&
 g= \left(       \begin{array}{cc}
         k_1^{-1} \D \mathbb{1}& 0   \\
             0 &   \mathbb{1}\\
           \end{array}
         \right)\ ,
         \quad  g^{-1}= \left(       \begin{array}{cc}
          k_1\D^{-1} \mathbb{1}& 0   \\
             0 &   \mathbb{1}\\
           \end{array}
         \right)\ ,
      \\
         &
   \tilde g= \left(       \begin{array}{cc}
          (1-\l^2) \mathbb{1}& -\l_0^{-1}\l\tilde \l  \mathbb{1}  \\
            -\l_0^{-1}\l\tilde \l  \mathbb{1} & (1-\l_0^{-2}\tilde \l^2)  \mathbb{1}\\
           \end{array}
         \right)\ ,
         \\ &
          \tilde g^{-1}= {k_1\ov \D} \left(       \begin{array}{cc}
          (1-\l_0^{-2}\tilde \l^2)  \mathbb{1}& \l_0^{-1}\l\tilde \l  \mathbb{1}  \\
            \l_0^{-1}\l\tilde \l  \mathbb{1} & (1-\l^2) \mathbb{1}\\
           \end{array}
         \right)\  .
    \end{split}
    \ee
Then the explicit form of the metric in the two-dimensional coupling space
spanned by $\l$ and $\tilde\l$ is found to be
\be
\begin{split}
& ds^2 = G_{11|11} d\L_{11}^2 + G_{21|21} d\L_{21}^2 + 2 G_{11|21} d\L_{11}d\L_{21}
\\
&\quad\ =  {k_1 \text{dimG}  \ov 2 \D^2} \left( (k_1-k_2 \tilde\l^2)d\l^2 + k_2 (1-\l^2)\l_0^{-2}d\tilde \l^2
+2 k_2 \l\tilde\l\ d\l d\tilde \l \right)\ .\
\label{lineelement}
\end{split}
\ee
Interestingly, this is, at least locally, an $AdS_2$ space since the corresponding Ricci scalar reads
$ R= -{4/\dim G}$. In addition, it can be shown that \eqn{lineelement} is invariant
under the transformation
\be
k_1\to -k_1\ , \qq \l\to {1\ov \l} \ ,\qq \tilde \l\to {\tilde \l \ov \l} \ .
\label{symmetry}
\ee
found in \cite{Georgiou:2018hpd} for the full effective action corresponding to \eqn{2couplingsaction}, as are the $\b$-functions \eqn{systrg} as well.

\subsection{Anomalous dimensions of the composite operators}

In order to compute the anomalous dimension of the bilinear current operators, we will follow the lines of \cite{Georgiou:2015nka,Sagkrioti:2018abh}.  For the general metric \eqn{metgggh}
the Cristoffel symbols were computed to be
\begin{align}
		\Gamma^{P_1P_2}_{M_1M_2|N_1N_2}
		=\d^{P_1}_{N_1}\d^{P_2}_{M_2}(\L g^{-1})_{M_1N_2}+\d^{P_1}_{M_1}\d^{P_2}_{N_2}(\L g^{-1})_{N_1M_2}\ .
\end{align}
The anomalous dimension matrix is taken from the work of \cite{Kutasov:1989dt}
\begin{equation}
	\gamma_{AB}{}^{CD}=\nabla_{AB}\b^{CD}+\nabla^{CD}\b_{AB}=\nabla_{AB}\b^{CD}+G_{AB|MN}G^{CD|PQ}\nabla_{PQ}\b^{MN}\ ,
\end{equation}
 where we have
used in here our double index notation so that
\be
\nabla_{AB}\b^{CD}=\partial_{AB}\b^{CD}+\Gamma^{CD}_{AB|MN}\b^{MN}\ ,\qq
 \partial_{AB}=\frac{\partial}{\partial\L_{AB}}\ .
\ee
It turns out that the non-zero components of the anomalous dimension tensor are
$\gamma_{ab}{}^{cd}$, $\g_{ab}{}^{c'd}$, $\gamma_{a'b}{}^{cd}$, $\gamma_{a'b}{}^{c'd}$, $\gamma_{ab'}{}^{cd'}$, $\gamma_{ab'}{}^{c'd'}$, $\gamma_{a'b'}{}^{cd'}$,  $\gamma_{a'b'}{}^{c'd'}$ and their explicit form can be found in the appendix \ref{self-mutapp}.
Due to the form of the interaction matrix $\L_{AB}$, we obtain the physical anomalous dimensions by "tracing" over the indices for each isotropic block of $\L$. We end up with
\begin{align}
	\begin{split}
		\gamma_{AB}{}^{cd}\d_{cd}=\left(\begin{matrix}
			\gamma_1\d_{ab}\quad 0\\ \gamma_2\d_{a'b}\quad 0
		\end{matrix}\right), \qq \gamma_{AB}{}^{c'd}\d_{c'd}=\left(\begin{matrix}
		\tilde{\gamma}_1\d_{ab}\quad 0\\ \tilde{\gamma}_2\d_{a'b}\quad 0
	\end{matrix}\right)\ ,
\end{split}
\end{align}
where
\be
 \label{eigenvalues1}
\begin{split}
&\gamma_1=-\frac{c_G}{\D^3}\left(2k_1^2\ f_1(\l)+k_2^2\ f_2(\l,\tilde{\l})-k_1k_2\ f_3(\l,\tilde{\l})   \right),\\
		&\gamma_2=\frac{c_G(1-\l)\l\tilde{\l}}{\D^4}\left(\sqrt{k_1^5k_2}\ f_4(\l,\tilde{\l})-\sqrt{k_1k_2^5}\ f_5(\l,\tilde{\l})+\sqrt{k_1^3k_2^3}\ f_6(\l,\tilde{\l})  \right)\
\end{split}
\ee
and	
\be
\begin{split}		
		&\tilde{\gamma}_1=\frac{c_G(1-\tilde{\l})\l\tilde{\l}}{\D^4}\left(\sqrt{k_1^5k_2}\ f_4(\l,\tilde{\l})-\sqrt{k_1k_2^5}\  f_5(\l,\tilde{\l})+\sqrt{k_1^3k_2^3}\ f_6(\l,\tilde{\l})  \right),\\
		&\tilde{\gamma}_2=\frac{c_G}{\D^3}\left( k_1^2\ f_7(\l,\tilde{\l})+k_2^2\ f_8(\tilde{\l})+k_1k_2\ f_9(\l,\tilde{\l}) \right)\  ,
		 \label{eigenvalues2}
	\end{split}
\ee
with
\begin{align}
\begin{split}
&f_1(\l)=(1-\l)^2\l\big(1-(1-\l)\l \big)\ ,\quad f_2(\l,\tilde{\l})=\l\tilde{\l}^4(2\tilde{\l}-3\l)\ ,
\\
&f_3(\l,\tilde{\l})=(1-\l)\tilde{\l}^2\big(3\l^3+(1-\tilde{\l})^2-5\l^2\tilde{\l}+\l(3+\tilde{\l}^2)  \big)\ ,
\\
&f_4(\l,\tilde{\l})=(1-\l)^2(1+\l)(2-\l+2\l^2-3\tilde{\l}(1+\l))\ ,
\\
&f_5(\l,\tilde{\l})=\tilde{\l}^4(3+3\l-2\tilde{\l})\ ,\quad f_6(\l,\tilde{\l})=\tilde{\l}^2(1-\l)\big(1+\tilde{\l}+\l(7+\l+\tilde{\l})\big)\ ,\\
&f_7(\l,\tilde{\l})=(1-\l)^2\big(\l^2-2\tilde{\l}(1+\l)(1+\l+\l^2)+3\tilde{\l}^2(1+\l)^2  \big)\ ,
\\ &f_8(\tilde{\l})=\tilde{\l}^4(3-2\tilde{\l})\ ,
\\ &f_9(\l,\tilde{\l})=\tilde{\l}^2(1-\l)\big(3+\l(3+\l(6+\l))-8\tilde{\l}-\l\tilde{\l}(8+5\l)+3\tilde{\l}^2(1+\l)  \big) \ .
\label{f-functions}
\end{split}
\end{align}
Hence, the anomalous dimension of $J_{1+}J_{1-}$ and $J_{2+}J_{1-}$ are
\begin{equation}
\label{g112}
\gamma_{J_{1+}J_{1-}}=\g_1+\tilde{\g}_1, \qq \gamma_{J_{2+}J_{1-}}=\g_2+\tilde{\g}_2\ .
\end{equation}

\subsubsection{Two limits and current anomalous dimensions}

In the $\tilde{\l}=0$ limit only the self-interaction $J_{1+}J_{1-}$ term in \eqn{2couplingsaction}
survives. Then \eqn{g112} simplifies to
	\begin{align}
	\begin{split}
	\gamma_{J_{1+}J_{1-}}=-\frac{2c_G}{k_1}\l\frac{1-\l(1-\l)}{(1-\l)(1+\l)^3},\qq
	\gamma_{J_{2+}J_{1-}}=\frac{c_G}{k_1}\frac{\l^2}{(1-\l)(1+\l)^3}\ .
	\end{split}
	\end{align}
In the above $\gamma_{J_{1+}J_{1-}}$ coincides with the anomalous dimension of $J_{1+}J_{1-}$ composite operator for the simply deformed model found in \cite{Georgiou:2015nka} (see also \eqn{jhjf}), while $\gamma_{J_{2+}J_{1-}}$ is the anomalous dimension of $J_{1-}$ (and $J_{1+}$ due to  isotropy) for the same model as in
\eqn{ansingg}. This should be the case since at this limit $J_{2+}$
is not interacting implying that $\gamma_{J_{2+}J_{1-}}=\gamma_{J_{1-}}=\g_{J_{1+}}$.

\no
In the $\l=0$ limit, only the self-interaction $J_{2+}J_{1-}$  \eqn{2couplingsaction} survives. Then, after also the rescaling
$\tilde{\l}\to \l_0\tilde{\l}$, we have that
		\begin{align}
		\begin{split}
		&\gamma_{J_{1+}J_{1-}}=\frac{c_G}{k_1}\tilde{\l}^2\frac{(1-\l_0\tilde{\l})^2}{(1-\tilde{\l}^2)^3}\ ,
		\\
		&\gamma_{J_{2+}J_{1-}}=\frac{c_G}{\sqrt{k_1k_2}}\tilde{\l}\frac{3(\l_0+\l_0^{-1})\tilde{\l}(1+\tilde{\l}^2)-2(1+4\tilde{\l}^2+\tilde{\l}^4)}{(1-\tilde{\l}^2)^3}\ .
		\end{split}
		\end{align}
As before, $\gamma_{J_{1+}J_{1-}}$ coincides with the anomalous dimension of $J_{1-}$ (since $J_{1+}$  is non-interacting),  while $\gamma_{J_{2+}J_{1-}}$ is the anomalous dimension
		of the composite operator found in \cite{Georgiou:2016zyo}.
The above derivation of anomalous dimensions of single and composite operators via a limiting procedure suggest that one may follow a similar procedure in more complicated models such as the ones in \cite{Georgiou:2018gpe}.

\section{Verification using perturbation theory}

In this section we proceed to verify perturbatively our previous exact results.
In order to do this, a set of relations are necessary for our calculation. All our perturbative calculations will be performed in Euclidean signature.

\no
The Callan-Symanzik equation implies that up to ${\cal O}(1/ k)$ the two-point function of an operator takes the form
\be
\label{Correl}
G_{ab}(x_1,x_2)=
G_0(k,\l)\frac{\d_{ab}}{x_{12}^{2\D}\bar{x}_{12}^{2\bar\Delta}}\Big(1+\g\ln\frac{\e^2}{|x_{12}|^2}+\dots \Big)\ ,
\ee
where $(\Delta,\bar\Delta)$ are the holomorphic and anti-holomorphic dimensions of the operator at the CFT point and $\gamma (\l)$ is the corresponding anomalous dimensions of ${\cal O}(1/k)$. The function $G_0(k,\l)$ is an overall normalization (which we should expand and keep up to ${\cal O}(1/k)$) and the short distance cutoff is  $\e$.
The indices $a,b$ take values in some generic irreducible representation of the group $G$.
Note that $\D$ and $\bar \D$ may depend on $k$, but in \eqn{Correl} only the $k$-independent
part is kept.

The $n$-point correlation function for a generic composite field $\Phi(x)$  is given by
\be
\langle \Phi(x_1)\cdots \Phi(x_n)\rangle=\frac{1}{Z}\int [D\Phi]\ \Phi(x_1)\cdots \Phi(x_n)e^{-S_k-{\l\ov \pi} \int d^2 z
J^a \bar J^a }
\ee
We will be interested in the anomalous dimensions of current composite operators. Therefore we will need
the  OPE for two holomorphic currents which reads
\be
J^a(z_1)J^b(z_2)=\frac{\d^{ab}}{z_{12}^2}+\frac{if^{abc}}{\sqrt{k}}\frac{J^c(z_2)}{z_{12}} \ ,
\ee
from which the two- and three-point functions
\be
\begin{split}
&\langle J^a(z_1)J^b(z_2)\rangle=\frac{\d^{ab}}{z_{12}^2} \ ,
\\
&\label{23pf}\langle J^a(z_1)J^b(z_2)J^c(z_3)\rangle=\frac{if^{abc}}{\sqrt{k}}\frac{1}{z_{12}z_{13}z_{23}}\ ,
\end{split}
\ee
follow. Higher correlators are evaluated using CFT Ward identities.
Finally, note that, one should be careful concerning the normal ordering of non-Abelian currents. In that respect we follow the prescription introduced in appendix A of \cite{BaisNorm}.
For two normal ordered product of any two operators operators $A$ and $B$ we will use the notation $(AB)$. For
the normal ordered product of more that two operators we will use the nested prescription, i.e. $(ABC)=(A(BC))$ etc.

\subsection{Chiral operators ${\cal O}^{(m,0)}$ to ${\cal O}(\l^2) $}

Clearly to ${\cal O}(\l)$  the anomalous dimension vanishes.
The first non-trivial contribution to the anomalous dimension of $\mathcal{O}^{(m,0)}$ comes from the two loop-contribution given by
\begin{align}
\begin{split}
&\langle\mathcal{O}^{(m,0)}(x_1)\mathcal{O}^{(m,0)}(x_2)\rangle_{\l^2}=\frac{\l^2}{2\p^2}\int d^2z_1d^2z_2\langle\bar{J}^{c_1}(\bar{z}_1)\bar{J}^{c_2}(\bar{z}_2)\rangle
\\
&\phantom{000000} \times d^{(m)}_{a_1\cdots a_m}d^{(m)}_{b_1\cdots b_m}
\langle (J^{a_1}\dots J^{a_m})(x_1)\,(J^{b_1}\dots J^{b_m})(x_2)\, J^{c_1}(z_1)J^{c_2}(z_2)\rangle\ .
\end{split}
\end{align}
Notice that all $k$-dependence comes entirely from the holomorphic correlation function.
Applying the Ward identity for the current at $z_1$ and suppressing momentarily the $d$-tensors the holomorphic part
of the correlator equals to
\begin{align}
\begin{split}
\label{JmultiJ}
&\quad\frac{m}{(z_1-x_1)^2}\langle\d^{c_1(a_1} (J^{a_2}\cdots J^{a_m)})(x_1)\, (J^{b_1}\dots J^{b_m})(x_2)J^{c_2}(z_2)\rangle \\
&+\frac{1}{\sqrt{k}}\frac{m}{(z_1-x_1)}\langle f^{ec_1(a_1} (J^{a_2}\dots J^{a_m)}J^e)(x_1)(J^{b_1}\dots J^{b_m})(x_2)J^{c_2}(z_2)\rangle\\
&+\frac{m}{(z_1-x_2)^2}\langle (J^{a_1}\dots J^{a_m})(x_1)\d^{c_1(b_1}(J^{b_2}\dots J^{b_m)})(x_2)J^{c_2}(z_2)\rangle\\
&+\frac{1}{\sqrt{k}}\frac{m}{
(z_1-x_2)}\langle (J^{a_1}\dots J^{a_m})(x_1)f^{ec_1(b_1}(J^{b_2}\dots J^{b_m)}J^e)(x_2)J^{c_2}(z_2)\rangle\ ,
\end{split}
\end{align}
where $m$ comes from the symmetrization conventions described at appendix \ref{group}.  Note that the contraction with $J^{c_2}(z_2)$ when combined with the anti-holomorphic contribution,
either vanishes  or gives rise to a bubble diagram. Performing the integral over $z_1$ for the first and third term,
we obtain terms proportional to $\d$-functions between internal and external points which are set to zero in our regularization
description \cite{Georgiou:2016iom} followed throughout this paper.\footnote{The issue of keeping the order of integration fixed \cite{Georgiou:2016iom}  and that of which delta functions are being kept is one and the same. One can keep all delta functions in which case the order of integration does not matter and the regulator $\epsilon$ is introduced whenever an integral diverges. We used the prescription of \cite{Georgiou:2016iom}  in order to minimize the number of integrals needed to be computed.}
Finally, applying the Ward identity for the current at $z_2$ and reinstating the $d$-tensors,
the holomorphic part of the correlator multiplied by $\d^{c_1c_2}$ (arising form the anti-holomorphic part of correlator)
 takes the form
\begin{align}
\begin{split}
\label{holcol}
&\quad\frac{m}{k}\frac{d^{(m)}_{a_1\dots a_m}d^{(m)}_{b_1\dots b_m}}{(z_2-x_1)(z_1-x_1)}\Big(f^{c_1a_1e}f^{c_1ef}\langle (J^{f}J^{a_1}\dots J^{a_m})(x_1)\, (J^{b_1}\dots J^{b_m})(x_2)\rangle\\
&\phantom{0000000000000000}+(m-1)f^{c_1a_1e}f^{c_1a_2f}\langle (J^{e}J^{f}J^{a_3}\dots J^{a_m})(x_1)\, (J^{b_1}\dots J^{b_m})(x_2)\rangle\Big)\\
&\quad +\frac{m^2}{k}\frac{d^{(m)}_{a_1\dots a_m}d^{(m)}_{b_1\dots b_m}f^{c_1a_1e}f^{c_1b_1f}}{(z_2-x_1)(z_1-x_2)}\langle (J^{e}J^{a_2}\dots J^{a_m})(x_1)\, (J^{f}J^{b_2}\dots J^{b_m})(x_2)\rangle\\
&\phantom{00000000000000000000}\qq +(x_1\leftrightarrow x_2)\ .
\end{split}
\end{align}
Having saturated the $1/k$ power, we only want the Abelian part of the remaining two-point function given by
\be
\label{multidelta}
\langle J^{a_1\dots a_m}(x_1)J^{b_1\dots b_m}(x_2)\rangle \big |_{\rm Abel}=
\frac{{m!}}{x_{12}^{2m}}\d^{a_1}_{(b_1}\d^{a_2}_{b_2}\dots\d^{a_m}_{b_m)}\ .
\ee
The parenthesis in the right hand side of \eqn{multidelta} denotes full symmetrisation of the indices $b_1,\dots, b_n$.
Plugging (\ref{multidelta}) in (\ref{holcol}) and using the identities (\ref{useidd}), (\ref{ffdid}) we conclude with
\be
\gamma_{\mathcal{O}^{(m,0)}}=0\ ,
\ee
up to ${\cal O}(\l^2)$,
in full agreement with (\ref{jjhjfh}).

\no
An important comment is in order.
The values of the operator anomalous dimension and the $\b$-function
at ${\cal O}(1/ k)$ do not depend on the regularization scheme,
but only on the very fact that the infinities are indeed absorbed to
${\cal O}(1/ k)$.
More quantitatively, looking at eqs. (3.14) and (3.15) of \cite{Georgiou:2016iom}  changing the finite part of  $Z_1$ and $Z$ is done in such a way that it does not affect not to ${\cal O}(1/ k)$ either $\gamma$ nor $\beta_\lambda$.
For the same reason the anomalous dimensions of all operators computed in this work are scheme independent to ${\cal O}(1/ k)$.
The scheme dependence will enter at the next order, namely at ${\cal O}(1/ k^2)$.

\subsection{Chiral operator ${\cal O}^{(2,0)}$ to ${\cal O}(\l^3) $}
Now consider the holomorphic operator $\mathcal{O}^{(2,0)}$ which is proportional to the holomorphic stress energy tensor $T$ (see, also the discussion around \eqn{tttpp}). Hence, subsequently we consider this operator for which the basic OPEs
are
\be
\begin{split}
& T(z)T(w)={c/2\ov (z-w)^2} + {2 T(w)\ov (z-w)^2} + {T'(w)\ov z-w}\ ,
\\
& J^a(z)T(w)=\frac{J^a(w)}{(z-w)^2}\ .
\end{split}
\ee
A set of four point functions necessary for our calculation, is
\be
\label{OOJJ}
\langle T(z_1)T(z_2)J^{a}(z_3)J^{b}(z_4)\rangle=\d^{ab}\Big(\frac{c/2}{z_{34}^2z_{12}^4}+\frac{1}{z_{12}^2z_{13}^2z_{24}^2}+\frac{1}{z_{12}^2z_{14}^2z_{23}^2}\Big)\ ,
\ee
where $\displaystyle c={2k\dim G\ov 2k +c_G}$ is the central charge of the WZW model theory
and
\be
\label{JJJO}
\langle J^{a_1}(z_1)J^{a_2}(z_2)J^{a_3}(z_3)T(z_4)\rangle=\frac{f^{a_1a_2a_3}}{\sqrt{k}}\Big(\frac{1}{z_{12}z_{24}^2z_{34}^2}-\frac{1}{z_{13}z_{24}^2z_{34}^2}+\frac{1}{z_{23}z_{14}^2z_{24}z_{34}}\Big)\ .
\ee
The first non-trivial loop contribution to the anomalous dimension comes at ${\cal O}(\l^2)$.
Since the energy momentum tensor is a quasi- instead of a primary field we repeat the computation of the previous subsection
which strictly speaking is valid for $m\geqslant 3$.
Then
\begin{align}
\langle T(x_1)T(x_2)\rangle_{\l^2}=\frac{\l^2}{2\p^2}\int d^2z_1d^2z_2\langle\bar{J}^{a}(\bar{z}_1)\bar{J}^b(\bar{z}_2)\rangle
 \langle T(x_1) T(x_2) J^{a_1}(z_1)J^{a_2}(z_2)\rangle\ .
\end{align}
Using (\ref{OOJJ}) we obtain for the connected part
\begin{align}
\begin{split}
&\langle T(x_1)T(x_2)\rangle_{\l^2}=\frac{\l^2}{2\p^2}\frac{\dim G}{x_{12}^2}\int d^2z_1d^2z_2\Big(\frac{1}{(z_1-x_1)^2(z_2-x_2)^2\bar{z}_{12}^2}\\
&\phantom{0000000000000000000000000000}+\frac{1}{(z_1-x_2)^2(z_2-x_1)^2\bar{z}_{12}^2}\Big)\ .
\\
\end{split}
\end{align}
Performing the $z_2$ integration, a delta-function $\d^{(2)}(z_2-x_i)$ with $i=1,2$ appears which is set to zero as explained
above. Then we arrive at
\be
\langle T(x_1) T(x_2)\rangle_{\l^2}=0\ .
\ee
Next, we continue to ${\cal O}(\l^3)$ with contribution
\begin{align}
\begin{split}
&\langle T(x_1)T(x_2)\rangle_{\l^3}=-\frac{\l^3}{6\p^3}\int d^2z_1d^2z_2d^2z_3\langle\bar{J}^{a_1}(\bar{z}_1)\bar{J}^{a_2}(\bar{z}_2)\bar{J}^{a_3}(\bar{z}_3)\rangle\times\\
&\phantom{000000000000000000000000000000}\langle T(x_1) T(x_2)J^{a_1}(z_1)J^{a_2}(z_2)J^{a_3}(z_3)\rangle\ .
\end{split}
\end{align}
The anti-holomorphic three point function is given by (\ref{23pf}).
Using for the Ward identity the operator at $z_1$, the holomorphic five-point function becomes
\begin{align}
\begin{split}
&\frac{if^{a_1a_2c}}{\sqrt{k}\ z_{12}}\langle J^c(z_2)J^{a_3}(z_3) T(x_1)T(x_2)\rangle+\frac{if^{a_1a_3c}}{\sqrt{k}}\frac{1}{z_{13}}\langle J^{a_2}(z_2)J^{c}(z_3)T(x_1)T(x_2)\rangle
\\
&+ \frac{1}{(z_1-x_1)^2}\langle J^{a_2}(z_2)J^{a_3}(z_3)J^{a_1}(x_1)T(x_2)\rangle+\frac{1}{(z_1-x_2)^2}\langle J^{a_2}(z_2)J^{a_3}(z_3)T(x_1)J^{a_1}(x_2)\rangle
\end{split}
\end{align}
 Utilizing once more  (\ref{OOJJ}) with (\ref{JJJO}) the three-loop contribution to the 2-point correlation function of $\mathcal{O}^{(2,0)}(x)$ takes the form
\begin{align}
\begin{split}
\label{Othreeloop}
&\langle T(x_1)T(x_2)\rangle_{\l^3}=\frac{c_G\dim G}{3\p^3x_{12}^2}\frac{\l^3}{k}\int \frac{d^2z_1d^2z_2d^2z_3}{\bar{z}_{12}\bar{z}_{13}\bar{z}_{23}}\times\\
&\phantom{00000000000000000}\Big\{\frac{1}{z_{12}(z_2-x_1)^2(z_3-x_2)^2}-\frac{1}{z_{13}(z_2-x_1)^2(z_3-x_2)^2}\\
&\phantom{0000000000000}+\frac{1}{z_{23}(z_1-x_1)^2(z_3-x_2)^2}-\frac{1}{(z_1-x_1)^2(z_2-x_1)(z_3-x_2)^2}\\
&\phantom{000000000000000000000}+\frac{x_{12}}{(z_1-x_1)^2(z_2-x_2)^2(z_3-x_1)(z_3-x_2)}\Big\}\ ,
\end{split}
\end{align}
where we have used the symmetry under $x_1\leftrightarrow x_2$ exchange to reduce the number of terms.
The detailed calculation of this
expression is given in appendix \ref{Appcorr} resulting into
\be
\langle T(x_1)T(x_2)\rangle_{\l^3}=0\ .
\ee
As expected, the  vanishing of the two-loop correlator and the duality symmetry of the action lock the all-loop correlation function of the $T\sim \mathcal{O}^{(2,0)}$ operator  to zero.

\subsection{The mixed operator ${\cal O}^{(2,1)}$ to ${\cal O}(\l^2) $}
The anomalous dimension of ${\cal O}^{(2,1)}$ is given by \eqn{jhjf}. For small values of $\l$ this becomes
\be
\label{anpert}
\g_ {{\cal O}^{(2,1)}} = -{2 c_G\ov k} (\l-3\l^2)+\mathcal{O}( \l^3)\ .
\ee
We would like to verify this against perturbation theory.

\no
For completeness we write down the OPE valid for our purposes
\begin{align}
\begin{split}
& J^a(z)(J^bJ^c)(w)=\frac{\d_{ab}}{(z-w)^2}J^c(w)+\frac{\d_{ac}}{(z-w)^2}J^b(w)+\frac{i}{\sqrt{k}}\frac{f_{abc}}{(z-w)^3}
\\	
&	\qq\quad +\frac{i}{\sqrt{k}}\frac{f_{abe}}{z-w}(J^eJ^c)(w)+\frac{i}{\sqrt{k}}\frac{f_{ace}}{z-w}(J^bJ^e)(w)-\frac{1}{k}\frac{f_{abe}f_{ecd}}{(z-w)^2}J^d(w)\ .
\end{split}
\end{align}
We need the tree-level contribution in order to properly normalize our results. We obtain that
\ba
\label{tree}
&& \langle\mathcal{O}^{(2,1)}(x_1)\mathcal{O}^{(2,1)}(x_2)\rangle_{\l^0}=d_{a_1a_2c_1}d_{b_1b_2c_2}\langle (J^{a_1}J^{a_2})(x_1)(J^{b_1}J^{b_2})(x_2)\rangle\langle \bar J^{c_1}(\bar{x}_1)J^{c_2}(\bar{x}_2)\rangle
\nonumber
\\
&& \qq\qq\qq\qq\phantom{xxxx}  = \frac{4(N^2-4)}{N}  \frac{\dim G}{x_{12}^4\bar{x}_{12}^2}\ ,
\ea
where we  used \eqn{multidelta}, \eqn{23pf} and \eqn{ddid}.

\no
The contribution to ${\cal O}(\l)$ is given by
\begin{align}
\begin{split}
\label{com3pf}
&\langle \mathcal{O}^{(2,1)}(x_1)\mathcal{O}^{(2,1)}(x_2)\rangle_{\l}=-\frac{\l}{\p}d_{a_1a_2c_1}d_{b_1b_2c_2}\int d^2z_1\langle\bar{J}^{c_1}(\bar{x}_1)\bar{J}^{c_2}(\bar{x}_2)\bar{J}^{d_1}(\bar{z}_1)\rangle
\\
&\phantom{0000000000000000000000000000}
\times \langle (J^{a_1}J^{a_2})(x_1)(J^{b_1}J^{b_2})(x_2)J^{d_1}(z_1)\rangle\\
\end{split}
\end{align}
Evaluating the holomorhic part separately we obtain that
\ba
\label{633}
&&\langle (J^{a_1}J^{a_2})(x_1)(J^{b_1}J^{b_2})(x_2)J^{d_1}(z_1)\rangle=\frac{2i}{\sqrt{k}(z_1-x_1)}f^{ed_1(a_1}\langle (J^{a_2)}J^{e})(x_1)(J^{b_1}J^{b_2})(x_2)\rangle
\nonumber\\
&&\phantom{00000xx0000000000000}+\frac{2i}{\sqrt{k}(z_1-x_2)}f^{ed_1(b_1}\langle (J^{b_2)}J^{e})(x_2)(J^{a_1}J^{a_2})(x_1)\rangle\ .
\ea
Putting everything together and using \eqn{multidelta} we obtain that
\ba
&& \langle\mathcal{O}^{(2,1)}(x_1)\mathcal{O}^{(2,1)}(x_2)\rangle_{\l} =\frac{4\l}{\p k}d_{a_1c_1a_2}d_{a_2c_2e}f_{ed_1a_1}\frac{f_{c_1c_2d_1}}{\bar x_{12}}\int \frac{d^2z_1}{(z_1-x_1)(\bar z_1-\bar x_1)(\bar z_1-\bar x_2)}
\nonumber
\\
&&\phantom{0000x00xx00}+\frac{4\l}{\p k}d_{b_1c_2b_2}d_{b_2c_1e}f_{ed_1b_1}\frac{f_{c_1c_2d_1}}{\bar x_{12}}\int \frac{d^2z_1}{(z_1-x_2)(\bar z_1-\bar x_1)(\bar z_1-\bar x_2)}\ .
\ea
Substituting \eqn{ffdd} and the integral \eqn{antholint} we arrive at
\be
\label{oneloop}
\langle \mathcal{O}^{(2,1)}(x_1)\mathcal{O}^{(2,1)}(x_2)\rangle=\frac{4(N^2-4)}{N}\frac{\dim G}{x_{12}^4\bar x_{12}^2}\Big(-\frac{2c_G}{k}\l\Big) \frac{1}{x_{12}^4\bar x_{12}^2}\ln\frac{\e^2}{|x_{12}|^2}\ .
\ee
We next consider the 2-loop contribution given by
\ba
&&\langle \mathcal{O}^{(2,1)}(x_1)\mathcal{O}^{(2,1)}(x_2)\rangle_{\l^2}=\frac{\l^2}{2\p^2}d_{a_1a_2c_1}d_{b_1b_2c_2}\int d^2z_1d^2z_2\langle \bar J^{c_1}(\bar x_1)\bar J^{c_2}(\bar x_2)\bar J^{d_1}(\bar z_1)\bar J^{d_2}(\bar z_2)\rangle
\nonumber
\\
&&\phantom{000000000000000000000000000000}
\times
\langle (J^{a_1}J^{a_2})(x_1)(J^{b_1}J^{b_2})(x_2)J^{d_1}(z_1)J^{d_2}(z_2)\rangle\ .
\ea
The anti-holomorphic part is just the 4-point function
\ba
&&\langle \bar J^{c_1}(\bar x_1)\bar J^{c_2}(\bar x_2)\bar J^{d_1}(\bar z_1)\bar J^{d_2}(\bar z_2)\rangle=\frac{\d^{d_1d_2}\d^{c_1c_2}}{\bar z_{12}^2\bar x_{12}^2}+\frac{\d^{d_1c_1}\d^{d_2c_2}}{(\bar z_1-\bar x_1)^2(\bar z_2-\bar x_2)^2}+\frac{\d^{d_1c_2}\d^{d_2c_1}}{(\bar z_1-\bar x_2)^2(\bar z_2-\bar x_1)^2}
\nonumber
\\
&&
\phantom{0000000000000000}-\frac{1}{k}\frac{1}{\bar x_{12}}\Big(\frac{f^{d_1d_2e}f^{c_1c_2e}}{\bar z_{12}(\bar z_2-\bar x_1)(\bar z_2-\bar x_2)}-\frac{f^{d_1c_1e}f^{d_2c_2e}}{(\bar z_1-\bar x_1)(\bar z_2-\bar x_1)(\bar z_2-\bar x_2)}
\\
&&\phantom{000000000000000000}+\frac{f^{d_1c_2e}f^{d_2c_1e}}{(\bar z_1-\bar x_2)(\bar z_2-\bar x_1)(\bar z_2-\bar x_2)}\Big)\ .
\nonumber
\ea
The holomorphic correlation function, having in mind that is multiplied with $d_{a_1a_2c_1}d_{b_1b_2c_2}$, is given by
\begin{align}
\begin{split}
\label{42loop}
\langle (J^{a_1}J^{a_2})(x_1)&(J^{b_1}J^{b_2})(x_2)J^{d_1}(z_1)J^{d_2}(z_2)\rangle={\frac{\d_{d_1d_2}}{z_{12}^2}\braket{(J^{a_1}J^{a_2})(x_1)(J^{b_1}J^{b_2})(x_2)}}
\\
	&{+\left(2+\frac{c_G}{2k}\right)\frac{\d_{d_1a_1}}{(z_1-x_1)^2}\braket{J^{a_2}(x_1)J^{d_2}(z_2)(J^{b_1}J^{b_2})(x_2)}}\\
&{+\left(2+\frac{c_G}{2k}\right)\frac{\d_{d_1b_1}}{(z_1-x_2)^2}\braket{(J^{a_1}J^{a_2})(x_1)J^{d_2}(z_2)J^{b_2}(x_2)}}\\
&{+\frac{i}{\sqrt{k}}\frac{f_{d_1a_1e}}{z_1-x_1}\braket{\big(J^eJ^{a_2}+J^{a_2}J^e\big)(x_1)J^{d_2}(z_2)(J^{b_1}J^{b_2})(x_2)}}\\
&{+\frac{i}{\sqrt{k}}\frac{f_{d_1b_1e}}{z_1-x_2}\braket{(J^{a_1}J^{a_2})(x_1)J^{d_2}(z_2)\big(J^eJ^{b_2}+J^{b_2}J^e\big)(x_2)}}\\
&{+\frac{i}{\sqrt{k}}\frac{f_{d_1d_2e}}{z_{12}}\braket{(J^{a_1}J^{a_2})(x_1)J^e(z_2)(J^{b_1}J^{b_2})(x_2)}}\ .
\end{split}
\end{align}
Note that one can easily evaluate
\be
\langle J^a(x_1)(J^bJ^c)(x_2)J^d(z)\rangle=\Big(2+\frac{c_G}{2k}\Big)\frac{\d^{d(b}\d^{c)a}}{(z-x_2)^2x_{12}^2}
\ee
and all three-point function of \eqn{42loop} are indices rearrangements of \eqn{633}.
To proceed, we are focused at the terms contributing to the anomalous dimension. Multiplying the holomorphic and anti-holomorphic parts there is a set of integrals  to evaluate.
Using \eqn{ffdd} and \eqn{ddid}, the integrals appearing are the following ones
\begin{align}
\begin{split}
\label{ints}
&\int\frac{d^2z_1d^2z_2}{(\bar z_1-\bar x_1)^2(z_1-x_1)(\bar z_2-\bar x_2)^2z_{12}}=-\frac{\p^2}{\bar x_{12}^2}\ln \frac{\e^2}{|x_{12}|^2}\ ,\\
&\int\frac{d^2z_1d^2z_2}{(\bar z_1-\bar x_1)^2(z_1-x_2)(\bar z_2-\bar x_2)z_{12}}=\frac{\p^2}{\bar x_{12}^2}\ln\frac{\e^2}{|x_{12}|^2}+\frac{\p^2}{\bar x_{12}^2} \ ,
\\
&\int \frac{d^2z_1d^2z_2}{(\bar z_1-\bar x_1)^2(z_2-x_1)(\bar z_2-\bar x_2)^2z_{12}}=-\frac{\p^2}{\bar x_{12}^2}\ln\frac{\e^2}{|x_{12}|^2}\ , \\
&\int\frac{d^2z_1d^2z_2}{(\bar z_1-\bar x_1)^2(z_2-x_2)(\bar z_2-\bar x_2)^2z_{12}}=\frac{\p^2}{\bar x_{12}^2}\ln\frac{\e^2}{|x_{12}|^2}+\frac{\p^2}{\bar x_{12}^2}\ , \\
&\int \frac{d^2z_1d^2z_2}{(\bar z_1-\bar x_1)(\bar z_2-\bar x_1)(\bar z_2-x_2)z_{12}^2}=\frac{\p^2}{\bar x_{12}}\ln\frac{\e^2}{|x_{12}|^2}\ , \\
&\int\frac{d^2z_1d^2z_2}{(\bar z_1-\bar x_2)(\bar z_2-\bar x_1)(\bar z_2-\bar x_2)z_{12}^2}=-\frac{\p^2}{\bar x_{12}}\ln\frac{\e^2}{|x_{12}|^2}\ , \\
&\int \frac{d^2z_1d^2z_2}{(z_1-x_1)^2(z_2-x_2)^2(\bar z_2-\bar x_1)(\bar z_2-\bar x_2)\bar z_{12}}=-\frac{2\p^2}{x_{12}^2\bar x_{12}}\ln\frac{\e^2}{|x_{12}|^2}-\frac{\p^2}{x_{12}^2\bar x_{12}}\ .
\end{split}
\end{align}
Having these, is an easy task to gather the terms contributing to the anomalous dimension. Performing the integrations using \eqn{ints}, we end up with
\be
\label{twoloop}
\langle \mathcal{O}^{(2,1)}(x_1)\mathcal{O}^{(2,1)}(x_2)\rangle_{\l^2}=\frac{1}{x_{12}^4\bar x_{12}^2}\frac{{4}(N^2-4)}{N}\frac{\dim G}{x_{12}^4\bar x_{12}^2}\Big(\frac{6c_G \l^2}{k}\Big) \ln\frac{\e^2}{|x_{12}|^2}\ .
\ee
Gathering \eqn{tree},\eqn{oneloop} and \eqn{twoloop} we get
\be
\langle \mathcal{O}^{(2,1)}(x_1)\mathcal{O}^{(2,1)}(x_2)\rangle=\frac{{4}(N^2-4)}{N} \frac{\dim G}{x_{12}^4\bar x_{12}^2}\Big(1 -\frac{2c_G}{k}(\l-3\l^2 )\ln\frac{\e^2}{|x_{12}|^2}\Big)+\cdots \ ,
\ee
where the ellipses denote terms contributing to order ${\cal O}(1/k)$ to the overall normalization or terms higher in the
small $\l$-expansion of the anomalous dimension in \eqn{Correl}. We conclude that the anomalous dimension is
indeed given by \eqn{anpert} up to ${\cal O}(\l^2)$.

\section{Discussion and future directions}

We demonstrated how the metric in the space of couplings and the all-loop effective action
allows for the calculation of the exact in the deformation parameters anomalous dimensions of composite operators in a wide class of integrable $\s$-models \cite{Sfetsos:2013wia,Georgiou:2016urf,Georgiou:2017jfi,Georgiou:2018hpd,Georgiou:2018gpe}. In our approach loop computations are completely avoided. The method relies on a generalization of the gauging procedure of \cite{Sfetsos:2013wia}. Specifically, it consists of adding to the gauged WZW model not only a gauged PCM term,
but also an irrelevant, generically Lorentz violating, term involving as many covariant derivatives as the number of currents building the composite operator whose dimension we seek. By fixing the gauge freedom  and integrating out the gauge fields, keeping however terms linear in the coupling of the composite irrelevant operator, one can obtain an effective action from which the $\beta$-functions of the model can de derived. Then, one uses \eqn{singleJ} to determine the anomalous dimension of the composite operator in terms of the $\beta$-functions and the metric in the space of couplings.

We considered deformations involving self- as well as mutually interacting current algebra theories.
We worked out the details for important classes of such operators.
In particular, we employed our method in order to calculate the anomalous dimensions of composite operators build from chiral and/or anti-chiral currents. As a first example we considered the operator built solely from an arbitrary number of same chirality currents in \eqn{om0} and in \eqn{om0-un} for the case of unequal levels. We showed that using the equations of motion, this operator is classically chiral in the $\l$-deformed theories even though the elementary currents are no longer such
away from the CFT point.
Surprisingly enough, their anomalous dimensions turn out to be zero to the leading order in the large $k$ expansion. This result allowed the preservation of the aforementioned chiral conservation laws up to ${\cal O}(1/k)$.
It will be interesting to investigate their fate to ${\cal O}(1/k^2)$ using in particular methods initiated in \cite{Goldschmidt:1980wq}.
In addition, we have also checked that the anomalous dimensions of composite operators which factorize into a chiral and an anti-chiral part is also vanishing to ${\cal O}(1/k)$.
Our last example concerned the fully symmetric operator composed from two chiral currents and one anti-chiral current. In
this its anomalous dimension turns out to be the same as that for the operator $J_+^a J_-^a$ driving the model off conformality.

\no
As a byproduct of our analyses, we have shown that the anomalous dimension of an operator that does not mix and to ${\cal O}(1/k)$ in the large $k$-expansion, vanishes if it does so, up to ${\cal O}(\l^2)$ in the small $\l$-expansion.

A number of other interesting questions remain to be addressed.
One of them is to calculate the anomalous dimensions of generic composite operators comprised of an arbitrary number of chiral  and anti-chiral currents or operators involving primary fields.
In considering this most general case one will have to deal with the serious problem of operator mixing.
Therefore it is imperative to search for a conceivable spin chain description. We expect that this spin chain will most likely be an integrable one at least for the cases where the underlying models are integrable as well.
One could also calculate the anomalous dimension of the single currents as well as composite operators made out of them
for the most general integrable models constructed in \cite{Georgiou:2018hpd} and \cite{Georgiou:2018gpe}. It is also possible since we have all the ingredients, albeit technically more difficult,
to compute the anomalous dimension of composite operators for the case of anisotropic couplings $\l_{ab}$.
Furthermore, it would certainly be very interesting to find the precise relation, if any, of these general integrable models  to those constructed recently in  \cite{Delduc:2018hty,Delduc:2019bcl} and see if our method can be used to derive the anomalous dimensions of composite operators in the latter models too.

\subsection*{Acknowledgments}

We would like to thank F. Delduc, M. Magro and K. Siampos for a useful discussions.\\
The work of G.G. on this project has received funding from the Hellenic Foundation for Research and Innovation
(HFRI) and the General Secretariat for Research and Technology (GSRT), under grant
agreement No 15425.\\
The research of E.S. is co-financed by Greece and the European Union (European Social Fund- ESF) through the Operational Programme ``Human Resources Development, Education and Lifelong Learning'' in the context of the project ``Strengthening Human Resources Research Potential via Doctorate Research'' (MIS-5000432), implemented by the State Scholarships Foundation (IKY).\\
K.S. would like to express a special thanks to the Mainz Institute for Theoretical Physics (MITP) of the Cluster of Excellence PRISMA+ (Project ID 39083149) for its hospitality and support during
the final stages of this research as well for the invitation to present these research results during the course of the program
``Holography, Generalized Geometry and Duality''.

\appendix

\section{The Zamolodchikov's metric for ${\cal O}^{(m,n)}$}
\label{zamome}

The purpose of this Appendix is to calculate the leading term in the large $k$-expansion of the  Zamolodchikov's metric for operators of the form \eqn{form-op}. We will do this  in the strict $k\to \infty$ limit in which the current algebra becomes Abelian.
The action arising from deforming a WZW action is \eqn{djkg11g} which we reproduce here for convenience in the Euclidean regime
\begin{align}
\begin{split}
&S=S_{WZW}-\frac{1}{\pi}\int d^2 z \left(\l\mathcal{O}+\tilde{\l}\mathcal{O}^{(m,n)} \right)\ ,
\\
&\mathcal{\tilde{O}}=J^a\J^a\ ,\quad  \mathcal{O}^{(m,n)} =S_{a_1\dots a_m;b_1\dots b_n}J^{a_1}\dots J^{a_m}\J^{b_1}\dots \J^{b_n} \ .
\label{general action}
\end{split}
\end{align}
We will now follow the lines of appendix A.2 in \cite{Georgiou:2015nka} in order to compute the exact in $\l$ and zeroth order in $\tilde{\l}$ Abelian part of the Zamolodchikov's metric in the coupling space of $\l$ and $\tilde \l$.
In what follows we will consider $m\geqslant n$. The resulting metric will be the same for $m< n$ as well.

\no
For the $G_{\l\l}=|x_{12}|^4 \braket{\mathcal{{O}}(x_1)\mathcal{{O}}(x_2)}$ part of the metric and to
${\cal O}(\tilde \l^0)$, the result is the same as in the simply deformed case computed in \cite{Georgiou:2015nka}.
The off-diagonal term $G_{\l\tilde{\l}}$ originating from the correlator $\braket{\mathcal{{O}}(x_1){\cal O}^{(m,n)}(x_2)}$ is zero to ${\cal O}(\tilde \l^0)$ in accordance with the assumption in \eqn{gggg}.
Thus, we only need to compute the Abelian part of $G_{\tilde{\l}\tilde{\l}}=x_{12}^{2m}\bar{x}_{12}^{2n} \braket{{\cal O}^{(m,n)}(x_1) {\cal O}^{(m,n)}(x_2)}$ term, exactly in $\l$ and to zeroth order in $\tilde{\l}$.

\no
Following the lines of \cite{Georgiou:2015nka}, we can expand this in powers of $\l$ as
\begin{equation}
\braket{{\cal O}^{(m,n)} {\cal O}^{(m,n)}}=G^{(0)}+\sum_{r=2,4,\dots }^{\infty}
\frac{\l^{r}}{\pi^{r}r!}G^{(r)}\ ,
\label{expansion}
\end{equation}
where clearly, since for $k\to \infty$ we have free oscillator contractions, only even terms contribute. We have that
\be
\begin{split}
&
G^{(0)}=S_{a_1\dots a_m;b_1\dots b_n}S_{c_1\dots c_m;d_1\dots d_n}\braket{J^{a_1}\dots J^{a_m}(x_1)J^{c_1}\dots J^{c_m}(x_2)}
\\
&
\qq\quad \times \braket{\J^{b_1}\dots\J^{b_n}(\bar{x}_1)\J^{d_1}\dots\J^{d_n}(\bar{x}_2)}
=\frac{m!n! S_{a_1\dots a_m;b_1\dots b_n}^2}{x_{12}^{2m}\bar{x}_{12}^{2n}}
\end{split}
\ee
and that
\ba
&&\hskip -.7 cm  G^{(r)}=\int d^2z_1\dots d^2z_r\ S_{a_1\dots a_m;b_1\dots b_n}s_{c_1\dots c_m;d_1\dots d_n}\braket{J^{a_1}\dots J^{a_m}(x_1)J^{e_1}(z_1)\dots J^{e_r}(z_r)J^{c_1}J^{c_m}(x_2)}
\nonumber\\
&&\phantom{0000000000}\times\braket{\J^{b_1}\dots \J^{b_n}(\bar{x}_1)\J^{e_1}(\z_1)\dots \J^{e_r}(\z_r)\J^{d_1}\dots \J^{d_n}(\bar{x}_2)}\ .
\ea
We can now find a recursive relation for $G^{(r)}$ by first recalling that
the points $z_1,\dots, z_k$ are internal, while $x_{1,2}$ are external ones and we must avoid disconnected and bubble diagrams.
Picking up the currents  $J^{e_1}(z_1)$ and $\J^{e_1}(\z_1)$ and keeping the above in mind we have contractions of the internal-internal and internal-external type.
The holomorphic $J^{e_1}$ can be contracted with any of the other $(r-1)$ internal currents $J^{e_i}$'s (with $j\neq 1$).
This should be  combined with the contraction of the
anti-holomorphic current $\J^{e_1}$  with one of the $(r-2)$ other internal $\J^{e_j}$'s (with $j\neq 1$, so that we avoid bubbles and disconnected diagrams) or with any of the $2n$ external
$\J^{b_i}$ and $\J^{d_i}$.
In addition, the holomorphic $J^{e_1}$ can be contracted with any of the $2 m$ external currents $J^{a_i}$ and $J^{c_i}$
and the result should be combined with the contraction of $\J^{e_1}$  with anyone of the $(r-1)$ other internal $\J^{e_i}$'s
(with $i\neq 1 $).
Thus, we have for $G^{(r)}$ the recursive relation
\begin{align}
\begin{split}
G^{(r)}&=\pi^2\left[(r-1)(r-2)+2n(r-1)+2m(r-1)\right]G^{(r-2)}
\\
&=\pi^2(r-1)\big(r+2(m+n-1)\big) G^{(r-2)}\ ,
\end{split}
\end{align}
where we used that $\displaystyle \int\frac{d^2z}{(z-x)^2(\z-\bar{y})^2}=\pi^2\d^{(2)}(x-y)$.
Solving it, we find that
\begin{equation}
G^{(r)}=\pi^{2k}\frac{(r-1)!!\big(r+2(m+n-1)\big)!!}{\big(2(m+n)-2\big)!!}G^{(0)}\ .
\end{equation}
Plugging this into (\ref{expansion}) and performing the sum we obtain the result
\begin{equation}
\braket{{\cal O}^{(m,n)} {\cal O}^{(m,n)}}
=\frac{m!n!S_{a_1\dots a_m;b_1\dots b_n}^2}{x_{12}^{2m}\bar{x}_{12}^{2n}(1-\l^2)^{m+n}}\ , \qq m\geqslant  n\ .
\end{equation}
Summarizing the above, the Abelian (for $k\to \infty$), exact in $\l$ but $\tilde{\l}$-independent components of the Zamolodchikov's metric are given by
\begin{align}
\label{memtrr}
\begin{split}
&G_{\l\l}={\dim G\ov (1-\l^2)^2}+\mathcal{O}(\tilde{\l})\ ,\quad G_{\l\tilde{\l}}=\mathcal{O}(\tilde{\l})\ ,
\\
& G_{\tilde{\l}\tilde{\l}}=\frac{{m!n!}S_{a_1\dots a_m;b_1\dots b_n}^2}{(1-\l^2)^{m+n}}+\mathcal{O}(\tilde{\l}).
\end{split}
\end{align}
Notice that in the special case of $m=2$, $n=0$ we should use that $S_{ab;0}=\d_{ab}$.


\section{Elements of $SU(N)$ group theory }
\label{group}

Here we use as references \cite{Invtens1, Invtens2, Invtens3}.
Consider a basis of $N\times N$ traceless matrices $\{t_a\}$,
$a=1,2,\dots , N^2-1$ and the normalization condition
\be
\Tr(t_a t_b)= \d_{ab}\ .
\ee
The multiplication law of two of them can be decomposed as
\be
t_a t_b = {\d_{ab}\ov N} \mathbb{I}_{N\times N} +\ha (if_{abc} + d_{abc}) t_c\ ,
\ee
where $f_{abc}$ is totally antisymmetric and $d_{abc}$ is symmetric and traceless. The coefficient
of the first term is dictated by the normalization condition. In our normalization the eigenvalue of the quadratic Casimir is
\be
c_G= 2 N \ .
\ee
In a given irreducible representation $R$ with elements $(t_a)_{\a\b}$, $\a,\b=1,2,\dots , \dim R$ the completeness relation
reads
\be
(t_a)_{\a\b}(t_a)_{\g\d} = \d_{\a\d} \d_{\b\g}-{1\ov N}\d_{\a\b}\d_{\g\d}\ .
\ee
The associativity property of matrix multiplication leads to the identities
\ba
&& f_{abe}f_{cfe} + f_{cae}f_{bfe} +  f_{bce}f_{afe} = 0\ ,
\nonumber\\
&&
d_{abe}f_{cfe} + d_{cae}f_{bfe} +  d_{bce}f_{afe} = 0\ ,
\label{ffsun}
\\
 && f_{abe}f_{cfe} = {4\ov N} (\d_{ac} \d_{bf} - \d_{bc} \d_{af}) +
d_{ace} d_{bfe} - d_{bce} d_{afe}\ .
\nonumber
\ea
From the last one by contracting with $\d_{bf}$ and relabeling
\be
\label{ddid}
d_{acd} d_{bcd} = 2 {N^2-4\ov N} \d_{ab}\ .
\ee
From \eqn{ffsun} by contracting and relabeling
\ba
\label{ffdd}
&& f_{eaf} f_{fbg} f_{gce} = -N f_{abc}\ ,
\nonumber\\
&& d_{eaf} f_{fbg} f_{gce} = -N d_{abc}\ ,
\nonumber\\
&& d_{eaf} d_{fbg} f_{gce} = {N^2-4\ov N} f_{abc}\ ,
\\
&& d_{eaf} d_{fbg} d_{gce} = {N^2-12\ov N} d_{abc}\ .
\nonumber
\ea
Symmetric tensors with more indices are computed recursively as
\be
d^{(m+1)}_{a_1a_2\dots a_{m+1}}
= d^{(m)}_{a ( a_1 a_2\dots a_{m-1}}d_{ a_m a_{m+1}) a}\ ,\qq m=3,4\dots \ ,
\ee
where in the symmetrization we include the appropriate weight factor.
For example
\be
d^{(4)}_{a_1 a_2 a_3 a_4} = {1\ov 3}(d_{ a a_1 a_2 }d_{a_3a_4 a}
+ d_{ a a_3 a_1 }d_{a_2 a_4 a} + d_{ a a_2 a_3 }d_{a_1 a_4 a})\ .
\label{dd44}
\ee

\no
The following is a useful identity
\be
f_{ab(a_1} d^{(m)}_{a_2a_3\dots a_m)b}=0 \ .
\label{useidd0}
\ee
It can be easily derived by recalling that $C^{(m)} = d_{a_1a_2\dots a_m} t^{a_1} t^{a_2} \dots t^{a_m} $ is a
Casimir operator and as such $[C^{(m)},t^a]=0, a=1,2,\dots \dim G$.
From that with appropriate contractions one obtains the  identity
 \be
 \label{useidd}
 d^{(m)}_{ab(a_1\dots a_{m-2}} f_{c)bd}f_{dae} =\D_m d_{c e a_1\dots a_{m-2}}\  , \qq  \D_m= -{c_G\ov m-1}\ ,\qq
m=2,3,\dots \ ,
 \ee
 as well as that
\be
\label{ffdid}
f_{dea}f_{db(a}d^{(m)}_{a_1\dots a_{m-2}c)b}=0\ .
\ee
Note that  $m=2$, we use the convention that $d_{ab}^{(2)}=\d_{ab}$.

\section{Supplement to the perturbative calculations}
\label{Appcorr}

We present the detailed calculation of the perturbative contributions to the anomalous dimensions discussed in the main text. For this, we first write down a list of integrals needed to be evaluated during our computation and use them when applicable.

\no
The basic technique for our computation is the use of Stokes theorem in two dimensions reading
\be
\int_{M} d^2x\,\partial_{\m}F^{\m}=\frac{i}{2}\int_{\partial M}\{d\bar zF^z-dzF^{\bar z}\}\ ,
\ee
where $M$ is the two-dimensional region and $\partial M$ the contour with positive rotational index when circles counterclockwise. Our integrating functions are not holomorphic so we cannot apply Cauchy's theorem but it can be easily
shown that the only parts contributing to the integrals appearing below are the  contours around the poles. In most cases we apply partial integration treating $z$ and $\bar z$ as independent variables, with a non-vanishing contribution coming from the superficial terms in general. In what follows $x_i$ denote external points while  $z_i$  internal, with $i=1,2$. The first set of integrals is the standard ones
\begin{align}
&\int \frac{d^2z}{(z-x_1)^2(\bar{z}-\bar{x}_2)^2}=\p \d^{(2)}(x_1-x_2)\ ,
\\
&\int\frac{d^2z}{(z-x_1)^2(\bar{z}-\bar{x}_2)}=-\frac{\p}{x_{12}}\ ,\qq \int\frac{d^2z}{(z-x_1)(\bar{z}-\bar{x}_2)^2}=\frac{\p}{\bar{x}_{12}}\ ,
\\
&\int  \frac{d^2z}{(z_1-x_1)(z_1-x_2)(\bar{z}_1-\bar{x}_2)}=\frac{\pi}{x_{12}}\ln\frac{\varepsilon^2}{|x_{12}|^2}\ ,
\\
&\label{antholint}\int  \frac{d^2z}{(z_1-x_1)(\bar{z}_1-\bar{x}_1)(\bar{z}_1-\bar{x}_2)}=-\frac{\pi}{\bar{x}_{12}}\ln\frac{\varepsilon^2}{|x_{12}|^2}\ .
\end{align}
Other integrals appearing in our computation of the anomalous dimension read
\begin{align}
&\label{A_1}\int\frac{d^2z}{(z-x_2)^2(\bar{z}-\bar{z}_1)(\bar{z}-\bar{z}_2)}=\frac{\p}{\bar{z}_{12}}\Big(\frac{1}{z_1-x_2}-\frac{1}{z_2-x_2}\Big)\ ,
\\
&\label{A_2}\int \frac{d^2z}{(z-z_1)(z-x_1)^2(\bar{z}-\bar{z}_1)^2}=\frac{\p}{\bar{z}_1-\bar{x}_1}\frac{1}{(z_1-x_1)^2}\ ,
\\
&\label{A_3}\int\frac{d^2z}{(z-x_2)(z-x_1)^2(\bar{z}-\bar{z}_1)^2}=\frac{\p}{x_{12}^2}\Big(\frac{1}{\bar{x}_2-\bar{z}_1}-\frac{1}{\bar{x}_1-\bar{z}_1}\Big)\ ,
\\
&\label{A_4}\int\frac{d^2z}{(z-x_2)(\bar{z}-\bar{x}_2)}-\int\frac{d^2z}{(z-x_2)(\bar{z}-\bar{x}_1)}=-\p\ln\frac{\e^2}{|x_{12}|^2}\ ,
\\
&\label{A_5}\int\frac{d^2z}{(z_1-z)(\bar{z}-\bar{z}_1)(\bar{z}-\bar{z}_2)}=\frac{\p}{\bar{z}_{12}}\ln\frac{\e^2}{|z_{12}|^2}\ ,
\\
&\label{A_6}\int\frac{d^2z}{(z-x_2)(\bar{z}-\bar{z}_1)(\bar{z}-\bar{z}_2)}=\frac{\p}{\bar{z}_{12}}\ln\frac{|z_2-x_2|^2}{|z_1-x_2|^2}\ ,
\\
&\label{A_7}\int\frac{d^2z}{(z-x_1)^2(\bar{z}-\bar{z}_{1})^2}\ln\frac{\e^2}{|z-z_1|^2}
=-\frac{\p}{|z_1-x_1|^2}\ ,
\end{align}
and
\begin{align}
\begin{split}
&\phantom{0000000}\label{A_8}\int\frac{d^2z}{(z-x_1)^2(\bar{z}-\bar{z}_{1})^2}\ln\frac{|z-x_2|^2}{|z_1-x_2|^2}
\\
&\qq\qq\qq =\frac{\p}{x_{12}}\Big(\frac{1}{\bar{x}_1-\bar{z}_1}-\frac{1}{\bar{x}_2-\bar{z}_1}\Big)
+\frac{\p}{(\bar{z}_1-\bar{x}_2)(z_1-x_1)}
\end{split}
\end{align}
and
\begin{align}
&\label{A_9}\int\frac{d^2z}{(z-x_1)(z-z_1)(\bar{z}-\bar{z}_{1})^2}=\frac{\p}{|z_1-x_1|^2}\ ,
\\
&\label{A_{10}}\int\frac{d^2z}{(z-x_1)(z-x_2)(\bar{z}-\bar{z}_{1})^2}=\frac{\p}{x_{12}}\Big(\frac{1}{\bar{x}_1-\bar{z}_1}-\frac{1}{\bar{x}_2-\bar{z}_1}\Big)\ ,
\\
&\label{A_{11}}\int  \frac{d^2z}{(z-x_2)^2(z-x_1)(\bar{z}-\bar{x}_1)}=-\frac{\p}{x_{12}^2}\ln\frac{\e^2}{|x_{12}|^2}-\frac{\p}{x_{12}^2}\ ,
\\
&\label{A_{12}}\int\frac{d^2z}{(z-x_2)^2(z-x_1)(\bar{z}-\bar{x}_2)}=\frac{\p}{x_{12}^2}\ln\frac{\e^2}{|x_{12}|^2}\ ,
\\
&\label{A_{13}}\int \frac{d^2z}{(z-x_2)^2(\bar{z}-\bar{z}_1)^2}\ln\frac{|z-x_2|^2}{|z_1-x_2|^2}=\frac{\p}{|z_1-x_2|^2}
\ .
\end{align}

\subsection{Holomorphic operator $\mathcal{O}^{(2,0)}$}
Going back to (\ref{Othreeloop}) there is a set of five integrals. We denote them by $I_1,\dots ,I_5$. Let us start with $I_1$ given by
\begin{align}
\begin{split}
\label{a}
&I_1=\int \frac{d^2z_1d^2z_2}{z_{12}(z_2-x_1)^2\bar{z}_{12}}\int\frac{d^2z_3}{(z_3-x_2)^2\bar{z}_{13}\bar{z}_{23}}\\
&\phantom{0}=\p\int \frac{d^2z_1}{z_1-x_2}\int \frac{d^2z_2}{z_{12}(z_2-x_1)^2\bar{z}_{12}^2}-\p\int \frac{d^2z_1d^2z_2}{z_{12}(z_2-x_1)^2(z_2-x_2)\bar{z}_{12}^2}\ ,
\\
\end{split}
\end{align}
where we used the splitting
\be
\frac{1}{\bar{z}_{13}\bar{z}_{23}}=\frac{1}{\bar{z}_{12}}\Big(\frac{1}{\bar{z}_3-\bar{z}_1}-\frac{1}{\bar{z}_3-\bar{z}_2}\Big)\ .
\ee
Applying the same splitting at the second integral, the first of the two terms appearing cancels the first integral leaving us with
\begin{align}
\begin{split}
&I_1=-\p\int\frac{d^2z_1}{z_1-x_2}\int \frac{d^2z_2}{(z_2-x_2)(z_2-x_1)^2\bar{z}_{12}^2}\ .
\\
\end{split}
\end{align}
The full splitting of the $z_2$ integral contains three terms
\be
\label{c}
\frac{1}{(z_2-x_2)(z_2-x_1)^2}=\frac{1}{x_{12}^2}\Big(\frac{1}{z_2-x_2}-\frac{1}{z_2-x_1}\Big)+\frac{1}{x_{12}}\frac{1}{(z_2-x_1)^2}\ ,
\ee
but the third vanishes in our renormalization scheme. Hence, we arrive at
\be
I_1=\frac{\p^2}{x_{12}^2}\Big(\int \frac{d^2z_1}{(z_1-x_2)(\bar{z}_1-\bar{x}_2)}-\int \frac{d^2z_1}{(z_1-x_2)(\bar{z}_1-\bar{x}_1)}\Big)\ .
\ee
Performing the integral we obtain that
\be
I_1=-\frac{\p^3}{x_{12}^2}\ln\frac{\e^2}{|x_{12}|^2}\ .
\ee
We proceed with $I_2$ given by
\be
I_2=\int \frac{d^2z_1d^2z_2}{(z_2-x_1)^2\bar{z}_{12}}\int\frac{d^2z_3}{z_{13}(z_3-x_2)^2\bar{z}_{13}\bar{z}_{23}}\ .
\ee
Using (\ref{c}), replacing $z_2$ with $z_3$ we obtain
\begin{align}
\begin{split}
&I_2=\int\frac{d^2z_1d^2z_2}{(z_2-x_1)^2\bar{z}_{12}}\Big(\frac{1}{(z_1-x_2)^2}\int\frac{d^2z_3}{z_{13}\bar{z}_{13}\bar{z}_{23}}\\
&\phantom{00}+\frac{1}{(z_1-x_2)^2}\int\frac{d^2z_3}{(z_3-x_2)\bar{z}_{13}\bar{z}_{23}}+\frac{1}{z_1-x_2}\int\frac{d^2z_3}{(z_3-x_2)^2\bar{z}_{13}\bar{z}_{23}}\Big)
\end{split}
\end{align}
and with the help of \eqn{A_5}, \eqn{A_6} and \eqn{A_1}, the integral $I_2$ becomes
\ba
\begin{split}
&I_2=\p\int \frac{d^2z_1}{(z_1-x_2)^2}\int \frac{d^2z_2}{(z_2-x_1)^2\bar{z}_{12}^2}\ln\frac{\e^2}{|z_{12}|^2}
\\
&
\qq +\p\int\frac{d^2z_1}{(z_1-x_2)^2}\int\frac{d^2z_2}{(z_2-x_1)^2\bar{z}_{12}^2}\ln\frac{|z_2-x_2|^2}{|z_1-x_2|^2}
\\
&\qq -\p\int\frac{d^2z_1}{z_1-x_2}\int\frac{d^2z_2}{(z_2-x_2)(z_2-x_1)^2\bar{z}_{12}^2}\ .
\end{split}
\ea
Now using \eqn{A_7}, \eqn{A_8} and \eqn{A_3} respectively, we arrive at
\begin{align}
\begin{split}
&I_2=-\p^2\int \frac{d^2z_1}{(z_1-x_2)^2(z_1-x_1)(\bar{z}_1-\bar{x}_1)}-\frac{\p^2}{x_{12}}\int\frac{d^2z_1}{(z_1-x_2)^2(\bar{z}_1-\bar{x}_1)}\\
&\phantom{0000}+\frac{\p^2}{x_{12}^2}\Big(\int\frac{d^2z_1}{(z_1-x_2)(\bar{z}_1-\bar{x}_2)}-\int\frac{d^2z_1}{(z_1-x_2)(\bar{z}_1-\bar{x}_1)}\Big)\\
&\phantom{0000}+\p^2\int\frac{d^2z_1}{(z_1-x_2)^2(z_1-x_1)(\bar{z}_1-\bar{x}_2)}
\end{split}
\end{align}
and finally from \eqn{A_{11}}, \eqn{A_4} and \eqn{A_{12}} respectively
\be
I_2=\frac{\p^3}{x_{12}^2}\ln\frac{\e^2}{|x_{12}|^2}\ .
\ee
Next we evaluate $I_3$ given by
\be
I_3=\int \frac{d^2z_1}{(z_1-x_1)^2}\int \frac{d^2z_2}{\bar{z}_{12}}\int\frac{d^2z_3}{z_{23}(z_3-x_2)^2\bar{z}_{13}\bar{z}_{23}}\ .
\ee
Applying (\ref{c}) we obtain
\begin{align}
\begin{split}
&I_3=\int \frac{d^2z_1}{(z_1-x_1)^2}\int\frac{d^2z_2}{\bar{z}_{12}}\Big\{\frac{1}{(z_2-x_2)^2}\int\frac{d^2z_3}{z_{23}\bar{z}_{13}\bar{z}_{23}}\\
&\phantom{0000}+\frac{1}{(z_2-x_2)^2}\int \frac{d^2z_3}{(z_3-x_2)\bar{z}_{13}\bar{z}_{23}}+\frac{1}{z_2-x_2}\int\frac{d^2z_3}{(z_3-x_2)^2\bar{z}_{13}\bar{z}_{23}}\Big\}\ .
\end{split}
\end{align}
Using \eqn{A_5} with $z_1\to z_2$, \eqn{A_6} and \eqn{A_1} respectively
\begin{align}
\begin{split}
&I_3=\int\frac{d^2z_1}{(z_1-x_1)^2}\Big\{-\int\frac{d^2z_2}{(z_2-x_2)^2\bar{z}_{12}}\ln\frac{\e^2}{|z_{12}|^2}\\
&\phantom{000}+\int\frac{d^2z_2}{(z_2-x_2)^2\bar{z}_{12}^2}\ln\frac{|z_2-x_2|^2}{|z_1-x_2|^2}+\frac{1}{z_1-x_2}\int\frac{d^2z_2}{(z_2-x_2)\bar{z}_{12}^2}\Big\}
\end{split}
\end{align}
and from  \eqn{A_3} with $x_1\to x_2$ and \eqn{A_{13}} respectively, a cancelation between the second and the third term of the resulting integral takes place leaving us  with
\be
I_3=\p^2\int\frac{d^2z_1}{(z_1-x_1)^2(z_1-x_2)(\bar{z}_1-\bar{x}_2)}=-\frac{\p^3}{x_{12}^2}\ln\frac{\e^2}{|x_{12}|^2}-\frac{\p^3}{x_{12}^2}\ ,
\ee
where we used  \eqn{A_{11}}.

\no
Next we calculate $I_4$ given by
\be
I_4=\int\frac{d^2z_1}{(z_1-x_1)^2}\int\frac{d^2z_2}{(z_2-x_1)\bar{z}_{12}}\int\frac{d^2z_3}{(z_3-x_2)^2\bar{z}_{13}\bar{z}_{23}}\ .
\ee
Using \eqn{A_1} we get
\begin{align}
\begin{split}
&I_4=\p\int\frac{d^2z_1}{(z_1-x_1)^2(z_1-x_2)}\int\frac{d^2z_2}{(z_2-x_1)\bar{z}_{12}^2}\\
&\phantom{000}-\p\int\frac{d^2z_1}{(z_1-x_1)^2}\int\frac{d^2z_2}{(z_2-x_1)(z_2-x_2)\bar{z}_{12}^2}
\end{split}
\end{align}
and a quick glimpse on \eqn{A_{10}} gives
\be
I_4=-\p^2\int\frac{d^2z_1}{(z_1-x_1)^2(z_1-x_2)(\bar{z}_1-\bar{x}_1)}-\frac{\p^2}{x_{12}}\int\frac{d^2z_1}{(z_1-x_1)^2(\bar{z}_1-\bar{x}_2)}\ .
\ee
Finally, with the help of \eqn{A_{12}} we arrive at
\be
I_4=-\frac{\p^3}{x_{12}^2}\ln\frac{\e^2}{|x_{12}|^2}+\frac{\p^3}{x_{12}^2}\ .
\ee
Our calculation is completed with $I_5$ where
\be
I_5=x_{12}\int\frac{d^2z_1}{(z_1-x_1)^2}\int\frac{d^2z_2}{\bar{z}_{12}(z_2-x_2)^2}\int\frac{d^2z_3}{(z_3-x_1)(z_3-x_2)\bar{z}_{13}\bar{z}_{23}}\ ,
\ee
which can be written as
\be
-x_{12}\int\frac{d^2z_1}{(z_1-x_1)^2}\int\frac{d^2z_2}{\bar{z}_{12}(z_2-x_2)^2}\Big\{\int\frac{d^2z_3}{(x_1-z_3)\bar{z}_{13}\bar{z}_{23}}+\int\frac{d^2z_3}{(z_3-x_2)\bar{z}_{13}\bar{z}_{23}}\Big\}
\ee
and exploiting \eqn{A_6} our previous relation becomes
\be
\begin{split}
&
{\p}\int\frac{d^2z_1}{(z_1-x_1)^2}\Big\{\int\frac{d^2z_2}{(z_2-x_2)^2\bar{z}_{12}^2}\ln\frac{|z_2-x_1|^2}{|z_1-x_1|^2}
\\
&\qq -\int\frac{d^2z_2}{(z_2-x_2)^2\bar{z}_{12}^2}\ln\frac{|z_2-x_2|^2}{|z_1-x_2|^2}\Big\}\ .
\end{split}
\ee
Using once more \eqn{A_8} and \eqn{A_{13}}
\begin{align}
\begin{split}
&I_5=\frac{\p^2}{x_{12}}\Big\{\int\frac{d^2z_1}{(z_1-x_1)^2(\bar{z}_1-\bar{x}_2)}+\int\frac{d^2z_1}{(z_1-x_1)^2(z_1-x_2)(\bar{z}_1-\bar{x}_1)}\\
&\phantom{0000000000000000000000}-\int\frac{d^2z_1}{(z_1-x_1)^2(z_1-x_2)(\bar{z}_1
-\bar{x}_2)}\Big\}\ ,
\end{split}
\end{align}
which gives
\be
I_5=\frac{2\p^3}{x_{12}^2}\ln\frac{\e^2}{|x_{12}|^2}\ .
\ee
Then gathering our results gives
\be
I_1+I_2+I_3+I_4+I_5=0\ ,
\ee
as stated in the main text.

\section{Supplement to section 5}
\label{self-mutapp}

In this Appendix, we present the anomalous dimension matrix for the model \cite{Georgiou:2018hpd} which contains
 both self and mutual current-current interactions.
We have that
 \be
\begin{split}
\gamma_{ab}{}^{cd}=\gamma_1\d_a^c\d_b^d\ , \quad \g_{ab}{}^{c'd}=\tilde{\g}_1\d_a^c\d_b^d\ ,\quad
\gamma_{a'b}{}^{cd}=\g_2\d_a^c\d_b^d\ , \quad \gamma_{a'b}{}^{c'd}=\tilde{\g}_2\d_a^c\d_b^d\ ,
\end{split}
\nonumber
\ee
 where the coefficients are given by \eqn{eigenvalues1} and \eqn{eigenvalues2}.
     Also
\begin{align*}
\begin{split}
&\gamma_{ab'}{}^{cd'}=\frac{c_G\l^2}{2\D^3}\left(-k_1^2f_{10}(\l)+k_2^2f_{11}(\l,\tilde{\l})+k_1k_2f_{12}(\l,\tilde{\l})  \right)\d_a^c\d_b^d\ ,
\\
&f_{10}(\l)=2\l(1-\l)^2\ ,\qq f_{11}(\l,\tilde{\l})=\l\tilde{\l}^4(\l-\tilde{\l})\ ,
\\\
&f_{12}(\l,\tilde{\l})=\tilde{\l}^2(1-\l)\big(2+3\l-\tilde{\l}(3+\l+\l^2)+\tilde{\l}^2(1+\l)  \big)\ ,
\end{split}
\end{align*}
\begin{align*}
\begin{split}
&\gamma_{ab'}{}^{c'd'}=\frac{c_G\l_0^{-1}\l\tilde{\l}}{2\D^3}\left(-k_1^2f_{13}(\l,\tilde{\l})+k_2^2f_{14}(\l,\tilde{\l})+k_1k_2f_{15}(\l,\tilde{\l})  \right)\d_a^c\d_b^d\ ,
\\
&f_{13}(\l,\tilde{\l})=\l\big(1-\l(3-\tilde{\l}^2)+\l^2(2-\tilde{\l})\big)+\tilde{\l}(1-\tilde{\l}),\qq f_{14}(\l,\tilde{\l})=f_{11}(\l,\tilde{\l})\ ,
\\
&f_{15}(\l,\tilde{\l})=\tilde{\l}^2\big(2+\l(1-2\l)-\tilde{\l}(3-\l-\l^3)+\tilde{\l}^2(1-\l^2)  \big)\ ,
\end{split}
\end{align*}
\begin{align*}
\begin{split}
&\gamma_{a'b'}{}^{cd'}=-\frac{c_G\l\tilde{\l}}{2\D^3}\left(\sqrt{k_1^3k_2}f_{16}(\l,\tilde{\l})+\sqrt{k_1k_2^3}f_{17}(\l,\tilde{\l})  \right)\d_a^c\d_b^d\ ,
\\
&f_{16}(\l,\tilde{\l})=(1-\l)^2\big(\l(1-\l)+\tilde{\l}(1+\l)(1+\l+\l^2)-\tilde{\l}^2(1+\l)^2  \big)\ ,
\\
&f_{17}(\l,\tilde{\l})=\tilde{\l}^2\big(-2+\l^2(1+\l^2)+\tilde{\l}(2-\l-\l^3)  \big)
\end{split}
\end{align*}
and
\begin{align*}
\begin{split}
&\gamma_{a'b'}{}^{c'd'}=\frac{c_G\tilde{\l}^2}{2\D^3}\left(k_2^2f_{18}(\l,\tilde{\l})-k_1k_2f_{19}(\l,\tilde{\l})\right)\d_a^c\d_b^d\\
&f_{18}(\l,\tilde{\l})=\tilde{\l}^2\big(2(1-\tilde{\l})-\l^3(\l-\tilde{\l})\big),\\
&f_{19}(\l,\tilde{\l})=\tilde{\l}\big(2(1-\tilde{\l})+\l^5\big)-\l^2\big((1-\tilde{\l})(2+3\tilde{\l})-\l(3-2\tilde{\l})+\l^2(1+\tilde{\l}^2)\big)
\end{split}
\end{align*}



\begin{thebibliography}{1}



 \bibitem{Sfetsos:2013wia}
  K.~Sfetsos, {\it Integrable interpolations: From exact CFTs to non-Abelian T-duals},\hfill\break
  Nucl. Phys. {\bf B880} (2014) 225, \href{http://arxiv.org/abs/arXiv:1312.4560}{arXiv:1312.4560 [hep-th]}.



\bibitem{Georgiou:2016urf}
  G.~Georgiou and K.~Sfetsos,
  {\it A new class of integrable deformations of CFTs},\\
   JHEP {\bf 1703} (2017) 083,
  \href{https://arxiv.org/abs/1612.05012}{arXiv:1612.05012 [hep-th]}.


  \bibitem{Georgiou:2017jfi}
  G.~Georgiou and K.~Sfetsos,
  {\it Integrable flows between exact CFTs},\\
  JHEP {\bf 1711}  (2017) 078,
   \href{https://arxiv.org/abs/1707.05149}{arXiv:1707.05149 [hep-th]}.

\bibitem{Georgiou:2018hpd}
  G.~Georgiou and K.~Sfetsos,
  {\it Novel all loop actions of interacting CFTs: Construction, integrability and RG flows},
  Nucl. Phys. {\bf B937} (2018) 371,\href{https://arxiv.org/abs/1809.03522}{arXiv:1809.03522 [hep-th]]}.

\bibitem{Georgiou:2018gpe}
  G.~Georgiou and K.~Sfetsos,
  {\it The most general $\lambda$-deformation of CFTs and integrability},
    JHEP {\bf 1903} (2019) 094,
  \href{https://arxiv.org/abs/1812.04033} {arXiv:1812.04033 [hep-th]}.

\bibitem{Driezen:2019ykp}
  S.~Driezen, A.~Sevrin and D.~C.~Thompson,
  {\it Integrable asymmetric $\lambda$-deformations},
  JHEP {\bf 1904}, 094 (2019)
  \href{https://arxiv.org/abs/1902.04142}{arXiv:1902.04142 [hep-th]}.

\bibitem{Georgiou:2016iom}
  G.~Georgiou, K.~Sfetsos and K.~Siampos,
  {\it All-loop correlators of integrable $\l$-deformed $\s$-models},
  Nucl.  Phys. {\bf B909} (2016) 360,
  \href{http://arxiv.org/abs/arXiv:1604.08212}{1604.08212 [hep-th].}

  \bibitem{Georgiou:2016zyo}
  G.~Georgiou, K.~Sfetsos and K.~Siampos,
  {\it $\lambda$-deformations of left-right asymmetric CFTs}, Nucl. Phys. {\bf B914} (2017) 623,
\href{https://arxiv.org/abs/1610.05314}{arXiv:1610.05314 [hep-th]}.



 \bibitem{Kutasov:1989aw}
  D.~Kutasov, {\it Duality Off the Critical Point in Two-dimensional Systems With Nonabelian Symmetries},
\href{http://www.sciencedirect.com/science/article/pii/0370269389913257}{Phys. Lett. {\bf B233} (1989) 369}.

    \bibitem{Itsios:2014lca}
  G.~Itsios, K.~Sfetsos and K.~Siampos,
  {\it The all-loop non-Abelian Thirring model and its RG flow},
  Phys.\ Lett.\  {\bf B733} (2014) 265,
  \href{http://arxiv.org/abs/1404.3748}{arXiv:1404.3748 [hep-th].}

  \bibitem{Sfetsos:2014jfa}
  K.~Sfetsos and K.~Siampos,
  {\it Gauged WZW-type theories and the all-loop anisotropic non-Abelian Thirring model},
  Nucl. Phys.  {\bf B885} (2014) 583,
  \href{http://arxiv.org/abs/arXiv:1405.7803}{arXiv:1405.7803 [hep-th].}

\bibitem{Georgiou:2015nka}
  G.~Georgiou, K.~Sfetsos and K.~Siampos,
  {\it All-loop anomalous dimensions in integrable $\lambda$-deformed $\sigma$-models},
  Nucl.\ Phys.\  {\bf B901} (2015) 40,
  \href{http://arxiv.org/abs/1509.02946}{arXiv:1509.02946 [hep-th].}


     \bibitem{Georgiou:2017aei}
  G.~Georgiou, E.~Sagkrioti, K.~Sfetsos and K.~Siampos,
  {\it Quantum aspects of doubly deformed CFTs},
Nucl. Phys. {\bf B919} (2017) 504,
 \href{https://arxiv.org/abs/1703.00462}
  {arXiv:1703.00462 [hep-th]}.

  \bibitem{Zamolodchikov:1986gt}
  A.B. Zamolodchikov,
 {\it Irreversibility of the Flux of the Renormalization Group in a 2D Field Theory},
\href{http://www.jetpletters.ac.ru/ps/1413/article_21504.shtml}{JETP Lett.  {\bf 43} (1986) 730}.

\bibitem{c-function:2018}
G. Georgiou, P. Panopoulos, E. Sagkrioti, K. Sfetsos, K. Siampos,
{\it The exact C-function in integrable $\l$-deformed theories},\\
Phys. Lett. {\bf B782} (2018) 613-18, \href{https://arxiv.org/abs/1805.03731}{arXiv:1805.03731 [hep-th]}.

\bibitem{Sagkrioti:2018abh}
  E.~Sagkrioti, K.~Sfetsos and K.~Siampos,
  {\it Weyl anomaly and the $C$-function in $\lambda$-deformed CFTs},
  Nucl. Phys. {\bf B938} (2019) 426,
  \href{https://arxiv.org/abs/1805.03731}{arXiv:1810.04189 [hep-th]}.

  \bibitem{Sfetsos:2014lla}
  K.~Sfetsos and K.~Siampos,
 {\it The anisotropic $\lambda$-deformed SU(2) model is integrable},
  Phys. Lett. {\bf B743} (2015) 160,
  \href{http://arxiv.org/abs/1412.5181}{arXiv:1412.5181 [hep-th]}.

 \bibitem{Sfetsos:2015nya}
  K.~Sfetsos, K.~Siampos and D.C.~Thompson,
 {\it Generalised integrable $\lambda$- and $\eta$-deformations and their relation},\hfill\break
  Nucl. Phys. {\bf B899} (2015) 489,
  \href{http://arxiv.org/abs/1506.05784}{arXiv:1506.05784 [hep-th]}.


   \bibitem{Kutasov:1989dt}
  D.~Kutasov,
  {\it String Theory and the Nonabelian Thirring Model},\\
 \href{http://www.sciencedirect.com/science/article/pii/0370269389912859}{Phys. Lett. {\bf B227} (1989) 68}.

  \bibitem{Gerganov:2000mt}
  B.~Gerganov, A.~LeClair and M.~Moriconi,
  {\it On the beta function for anisotropic current interactions in 2-D},
  Phys. Rev. Lett. {\bf 86} (2001) 4753,
 \href{http://arxiv.org/abs/hep-th/0011189}{hep-th/0011189}.

 \bibitem{LeClair:2001yp}
  A.~LeClair,
  {\it Chiral stabilization of the renormalization group for flavor and color anisotropic current interactions},
  Phys.\ Lett.\ {\bf B519} (2001) 183,
  \href{https://arxiv.org/abs/hep-th/0105092v2}{hep-th/0105092}.

  \bibitem{Appadu:2015nfa}
  C. Appadu and T.J. Hollowood,
  {\it Beta function of k deformed ${\text AdS}_{5} \times S^5$ string theory},
  JHEP {\bf 1511} (2015) 095,
  \href{http://arxiv.org/abs/arXiv:1507.05420}{arXiv:1507.05420 [hep-th].}

\bibitem{Georgiou:2017oly}
  G.~Georgiou, K.~Sfetsos and K.~Siampos,
  {\it Double and cyclic $\lambda$-deformations and their canonical equivalents},
  Phys. Lett. {\bf B771}  (2017) 576,
   \href{https://arxiv.org/abs/1704.07834}{arXiv:1704.07834 [hep-th]}.

   \bibitem{Sagkrioti:2018rwg}
  E.~Sagkrioti, K.~Sfetsos and K.~Siampos,
  {\it RG flows for $\lambda$-deformed CFTs},\\
  Nucl.\ Phys.\ {\bf B930} (2018) 499,
    \href{https://arxiv.org/abs/1801.10174}{arXiv:1801.10174 [hep-th]}.

    \bibitem{honer}
 G.~Ecker and J.~Honerkamp,
 {\it Application of invariant renormalization to the nonlinear chiral invariant
 pion Lagrangian in the one-loop approximation},\hfill\break
 \href{http://www.sciencedirect.com/science/article/pii/0550321371904688}{Nucl. Phys. {\bf B35} (1971) 481}.\hfill\break
J.~Honerkamp,
 {\it Chiral multiloops},
\href{http://www.sciencedirect.com/science/article/pii/0550321372902994}{Nucl. Phys. {\bf B36} (1972) 130}. \hfill\break
  D.~Friedan,
  {\it Nonlinear Models in Two Epsilon Dimensions},\hfill\break
  \href{http://journals.aps.org/prl/abstract/10.1103/PhysRevLett.45.1057}{Phys. Rev. Lett. {\bf 45} (1980) 1057}
 and {\it Nonlinear Models in Two + Epsilon Dimensions},
  \href{http://www.sciencedirect.com/science/article/pii/0003491685903847}{Annals Phys. {\bf 163} (1985) 318}.



\bibitem{Balog:1993es}
  J.~Balog, P.~Forgacs, Z.~Horvath and L.~Palla,
  {\it A New family of $SU(2)$ symmetric integrable sigma models,}
  Phys. Lett. {\bf B324} (1994) 403,
  \href{http://arxiv.org/abs/hep-th/9307030}{hep-th/9307030.}

 \bibitem{Sfetsos:2017sep}
  K.~Sfetsos and K.~Siampos,
  {\it Integrable deformations of the $G_{k_1} \times G_{k_2}/G_{k_1+k_2}$ coset CFTs},
  Nucl. Phys. {\bf B927} (2018) 124,
  \href{https://arxiv.org/abs/1710.02515}{arXiv:1710.02515  [hep-th]}.

 \bibitem{Hollowood:2014rla}
  T.J.~Hollowood, J.L.~Miramontes and D.M.~Schmidtt,
 {\it Integrable Deformations of Strings on Symmetric Spaces},
  JHEP {\bf 1411} (2014) 009,
  \href{http://arxiv.org/abs/1407.2840}{arXiv:1407.2840 [hep-th]}.



\bibitem{Hollowood:2014qma}
  T.J.~Hollowood, J.L.~Miramontes and D.~Schmidtt,
{\it An Integrable Deformation of the $AdS_5 \times S^5$ Superstring},
J.\ Phys.\ {\bf A47} (2014) 49,  495402,
 \href{http://arxiv.org/abs/1409.1538}{arXiv:1409.1538 [hep-th]}.

 \bibitem{Sfetsos:2014cea}
  K.~Sfetsos and D.C.~Thompson,
  {\it Spacetimes for $\lambda$-deformations},\hfill\break
  JHEP {\bf 1412} (2014) 164,
  \href{http://arxiv.org/abs/1410.1886}{arXiv:1410.1886 [hep-th]}.



  \bibitem{Demulder:2015lva}
  S.~Demulder, K.~Sfetsos and D.C.~Thompson,
  {\it Integrable $\lambda$-deformations: Squashing Coset CFTs and $AdS_5\times S^5$},
  JHEP {\bf 07} (2015) 019,
  \href{http://arxiv.org/abs/1504.02781}{arXiv:1504.02781 [hep-th]}.


\bibitem{Borsato:2016zcf}
  R.~Borsato, A.A.~Tseytlin and L.~Wulff,
  {\it Supergravity background of $\lambda$-deformed model for AdS$_2 \times$  S$^2$ supercoset},
  Nucl. Phys. {\bf B905} (2016) 264,
  \href{http://arxiv.org/abs/1601.08192}{arXiv:1601.08192 [hep-th]}.

\bibitem{Chervonyi:2016ajp}
  Y.~Chervonyi and O.~Lunin,
  {\it Supergravity background of the $\lambda$-deformed $\text{AdS}_3 \times S^3$ supercoset},
  Nucl. Phys. {\bf B910} (2016) 685,
  \href{https://arxiv.org/abs/1606.00394}{arXiv:1606.00394 [hep-th]}.

\bibitem{Borsato:2016ose}
  R.~Borsato and L.~Wulff,
  {\it Target space supergeometry of $\eta$ and $\lambda$-deformed strings},
  JHEP {\bf 1610} (2016) 045,
   \href{https://arxiv.org/abs/1608.03570}{arXiv:1608.03570 [hep-th]}.

 \bibitem{Klimcik:2002zj}
C. Klim\v c\'\i k,
  {\it YB sigma models and dS/AdS T-duality},\hfill\break
  JHEP {\bf 0212} (2002) 051,
\href{http://arxiv.org/abs/hep-th/0210095}{hep-th/0210095}.

\bibitem{Klimcik:2008eq}
  C. Klim\v c\'\i k,
  {\it On integrability of the YB sigma-model},\hfill\break
  J. Math. Phys. {\bf 50} (2009) 043508,
  \href{http://arxiv.org/abs/0802.3518}{arXiv:0802.3518 [hep-th]}.

 \bibitem{Klimcik:2014}
  C. Klim\v c\'\i k,
  {\it Integrability of the bi-Yang--Baxter sigma-model},
  Letters in Mathematical Physics {\bf 104} (2014) 1095,
    \href{http://arxiv.org/abs/1402.2105}{arXiv:1402.2105 [math-ph]}.

   \bibitem{Delduc:2013fga}
  F.~Delduc, M.~Magro and B.~Vicedo,
{\it On classical $q$-deformations of integrable sigma-models},
  JHEP {\bf 1311} (2013) 192,
   \href{http://arxiv.org/abs/1308.3581}{arXiv:1308.3581 [hep-th]}.

\bibitem{Delduc:2013qra}
  F.~Delduc, M.~Magro and B.~Vicedo,
{\it An integrable deformation of the $AdS_5 \times S^5$ superstring action},
  Phys. Rev. Lett. {\bf 112}, 051601,
     \href{http://arxiv.org/abs/1309.5850}{arXiv:1309.5850 [hep-th].}

\bibitem{Arutyunov:2013ega}
  G.~Arutyunov, R.~Borsato and S.~Frolov,
  {\it S-matrix for strings on $\eta$-deformed $AdS_{5} \times S^5$},
  JHEP {\bf 1404} (2014) 002,
 \href{http://arxiv.org/abs/1312.3542}{arXiv:1312.3542 [hep-th].}


  \bibitem{KS95a}{C. Klim\v c\'\i k and P. \v Severa, {\it Dual non-Abelian duality and the Drinfeld double},\\
Phys. Lett. {\bf B351}
(1995) 455, \href{http://arxiv.org/abs/hep-th/9502122}{hep-th/9502122}}.

\bibitem{Sfetsos:1999zm}
  K.~Sfetsos,
  {\it Duality invariant class of two-dimensional field theories},\hfill\break
  Nucl. Phys. {\bf B561} (1999) 316,
  \href{https://arxiv.org/abs/hep-th/9904188}{[hep-th/9904188]}.

\bibitem{Vicedo:2015pna}
  B.~Vicedo,
  {\it Deformed integrable $\sigma$-models, classical $R$-matrices and classical exchange algebra on Drinfel'd doubles},
  \hfill\break
  J. Phys. A: Math. Theor. {\bf 48} (2015) 355203,
 \href{http://arxiv.org/abs/1504.06303}{arXiv:1504.06303 [hep-th].}

\bibitem{Hoare:2015gda}
  B.~Hoare and A.A.~Tseytlin,
  {\it On integrable deformations of superstring sigma models related to $AdS_n \times S^n$ supercosets},\hfill\break
  {Nucl. Phys. {\bf B897} (2015) 448},
  \href{http://arxiv.org/abs/1504.07213}{arXiv:1504.07213 [hep-th].}














   \bibitem{Klimcik:2015gba}
C. Klim\v c\'\i k,
  {\it $\eta$ and $\lambda$ deformations as ${\cal E}$-models},\hfill\break
   Nucl. Phys. {\bf B900} (2015) 259,
  \href{http://arxiv.org/abs/1508.05832}{arXiv:1508.05832 [hep-th].}


\bibitem{Klimcik:2016rov}
C. Klim\v c\'\i k,
  {\it Poisson--Lie T-duals of the bi-Yang--Baxter models},\hfill\break
  Phys. Lett.  {\bf B760} (2016) 345,
  \href{https://arxiv.org/abs/1606.03016}{arXiv:1606.03016 [hep-th]}.

\bibitem{Hoare:2018ebg}
  B.~Hoare and F.K.~Seibold,
 {\it Poisson-Lie duals of the $\eta$-deformed $\mathrm{AdS}_2 \times \mathrm{S}^2 \times \mathrm{T}^6$ superstring},
  JHEP {\bf 1808} (2018) 107,
 \href{https://arxiv.org/abs/1807.04608}{arXiv:1807.04608 [hep-th]}.

\bibitem{Lunin:2018vsn}
  O.~Lunin and W.~Tian,
  {\it Scalar fields on $\lambda$-deformed cosets},\hfill\\
  Nucl. Phys. {\bf B938} (2019) 671,
   \href{https://arxiv.org/abs/1808.02971}{arXiv:1808.02971 [hep-th]}.

\bibitem{Schmidtt:2017ngw}
  D.M.~Schmidtt,
  {\it Integrable Lambda Models And Chern-Simons Theories},\hfill\break
  JHEP {\bf 1705} (2017) 012,
   \href{https://arxiv.org/abs/1808.05994}{arXiv:1701.04138 [hep-th]}
 \hfill\break
 and {\it Lambda Models From Chern-Simons Theories},\\
 JHEP {\bf 1811} (2018) 111,
    \href{https://arxiv.org/abs/1808.05994}{arXiv:1808.05994 [hep-th]}.

\bibitem{Driezen:2018glg}
  S.~Driezen, A.~Sevrin and D. C.~Thompson,
  {\it D-branes in $\lambda$-deformations},\hfill\break
  JHEP {\bf 1809} (2018) 015,
 \href{https://arxiv.org/abs/1806.10712}{arXiv:1806.10712 [hep-th]}.



\bibitem{Witten:1983ar}
  E.~Witten,
  {\it Nonabelian Bosonization in Two-Dimensions},\hfill\break
  \href{https://link.springer.com/article/10.1007\%2FBF01215276}{Commun.\ Math.\ Phys.\  {\bf 92} (1984) 455.}

\bibitem{Polyakov:1975rr}
  A.M.~Polyakov,
  {\it Interaction of Goldstone Particles in Two-Dimensions. Applications to Ferromagnets and Massive Yang-Mills Fields},
 \href{http://www.sciencedirect.com/science/article/pii/0370269375901616}{Phys. Lett. {\bf B59} (1975) 79}.\hfill\break
  K.~Pohlmeyer,
  {\it Integrable Hamiltonian Systems and Interactions Through Quadratic Constraints},
  \href{http://link.springer.com/article/10.1007%2FBF01609119
}{Commun. Math. Phys.  {\bf 46} (1976) 207}.\hfill\break
  M.~Luscher,
  {\it Quantum Nonlocal Charges and Absence of Particle Production in the Two-Dimensional Nonlinear Sigma Model},
 \href{http://www.sciencedirect.com/science/article/pii/0550321378902110}{Nucl. Phys. {\bf B135} (1978) 1}.\hfill\break
  M.~Luscher and K.~Pohlmeyer,
  {\it Scattering of Massless Lumps and Nonlocal Charges in the Two-Dimensional Classical Nonlinear Sigma Model},
  \href{http://www.sciencedirect.com/science/article/pii/0550321378900494}{Nucl. Phys. {\bf B137} (1978) 46}.

\bibitem{gwzwac}
  D.~Karabali, Q.H.~Park, H.J.~Schnitzer and Z.~Yang,
  {\it A GKO Construction Based on a Path Integral Formulation of Gauged Wess-Zumino-Witten Actions},\hfill\break
  \href{http://www.sciencedirect.com/science/article/pii/0370269389911209}{Phys. Lett.  {\bf B216} (1989) 307}.\hfill\\
  D.~Karabali and H.J.~Schnitzer,
  {\it BRST Quantization Of The Gauged WZW Action And Coset Conformal Field Theories},
  \href{http://www.sciencedirect.com/science/article/pii/055032139090075O}{Nucl. Phys. {\bf B329} (1990) 649}.\hfill\\
  K.~Gawedzki and A.~Kupiainen,
  {\it $G/H$ Conformal Field Theory from Gauged WZW Model},
  \href{http://www.sciencedirect.com/science/article/pii/0370269388910817}{Phys. Lett. {\bf B215} (1988) 119}.

\bibitem{BaisNorm}
  F.A.~Bais, P.~Bouwknegt, M.~Surridge and K.~Schoutens,
  {\it Extensions Of The Virasoro Algebra Constructed From Kac-Moody Algebras
  Using Higher Order Casimir Invariants},
 \href{https://www.sciencedirect.com/science/article/pii/0550321388906311?via\%3Dihub}
  {Nucl. Phys. {\bf B304} (1988) 348}.

\bibitem{Goldschmidt:1980wq}
  Y.~Y.~Goldschmidt and E.~Witten,
  {\it Conservation Laws in Some Two-dimensional Models},
  \href{https://www.sciencedirect.com/science/article/abs/pii/0370269380910047?via\%3Dihub}{Phys. Lett. {\bf B91} (1980) 392}

\bibitem{Delduc:2018hty}
  F.~Delduc, S.~Lacroix, M.~Magro and B.~Vicedo, {\it Integrable coupled sigma-models},
   Phys. Rev. Lett. {\bf 122} (2019) no.4, 041601,
  \href{https://arxiv.org/abs/1811.12316}{ arXiv:1811.12316 [hep-th]}.

\bibitem{Delduc:2019bcl}
  F.~Delduc, S.~Lacroix, M.~Magro and B.~Vicedo,
  {\it Assembling integrable $\sigma$-models as affine Gaudin models},
  JHEP {\bf 1906} (2019) 017,
   \href{https://arxiv.org/abs/1903.00368}{  arXiv:1903.00368 [hep-th]}.

 \bibitem{Invtens1}
J.A. de Azcarraga, A.J. Macfarlane, A.J. Mountain, J.C. Perez Bueno,
{\it Invariant tensors for simple groups}, Nucl.Phys. {\bf B510} (1998) 657-687,
\href{http://arxiv.org/abs/physics/9706006}{physics/9706006}.

\bibitem{Invtens2}
L.M. Kaplan and M. Resnikoff, {\it Matrix Products
and the Explicit 3, 6, 9, and 12-j Coefficients of the Regular Representation of $SU(n)$},
\href{https://aip.scitation.org/doi/10.1063/1.1705141}{Math. Phys. {\bf 8} (1967) 2194}.

\bibitem{Invtens3}
A.J. MacFarlane, A. Sudbery, and P.H. Weisz,
{\it On Gell-Mann's $\l$-Matrices, d- and f-tensors, octets, and parametrizations of $SU(3)$},
\hfill\\
\href{https://link.springer.com/article/10.1007/BF01654302}{Comm. Math. Phys. Volume {\bf 11} (1968) 77-90}.

  \end{thebibliography}
\end{document}